\documentclass[12pt]{article} 
\usepackage{epsf}  
\usepackage{float}  
\begin{document}
\baselineskip 14pt plus 2pt minus 2pt
\newcommand{\be}{\begin{equation}}
\newcommand{\ee}{\end{equation}}
\newcommand{\bea}{\begin{eqnarray}}
\newcommand{\eea}{\end{eqnarray}}
\newcommand{\no}{\nonumber}
\newcommand{\w}{\omega}
\newcommand{\de}{\Delta}
\newcommand{\gpnd}{g_{\pi N \Delta}}
\newcommand{\beq}{\begin{equation}}
\newcommand{\eeq}{\end{equation}}
\newcommand{\beqa}{\begin{eqnarray}}
\newcommand{\eeqa}{\end{eqnarray}}
\newcommand{\dfrac}{\displaystyle \frac}
\renewcommand{\thefootnote}{\#\arabic{footnote}}
\newcommand{\ve}{\varepsilon}
\newcommand{\krig}[1]{\stackrel{\circ}{#1}}
\newcommand{\barr}[1]{\not\mathrel #1}
\newcommand{\vs}{\vspace{-0.2cm}}
\def\ve{\varepsilon}
\def\lanl{\langle}
\def\ranl{\rangle}

\begin{titlepage}
 
\hfill {\tiny {\bf FZJ-IKP(TH)-2000-13}}

\vspace{1.5cm}

\begin{center}

{\large  \bf {
Pion--nucleon scattering in an effective chiral field theory\\[0.2em]
with explicit spin--3/2 fields\footnote{Work supported in part by a NATO
  research grant under no. BOT99/001.}}}

\vspace{1.2cm}
                              
{\large Nadia Fettes\footnote{email: N.Fettes@fz-juelich.de},
Ulf-G. Mei\ss ner\footnote{email: Ulf-G.Meissner@fz-juelich.de}
}

\vspace{1.0cm}

{\it Forschungszentrum J\"ulich, Institut f\"ur Kernphysik (Theorie)}

{\it D--52425 J\"ulich, Germany}\\

\end{center}

\vspace{0.8cm}

\begin{abstract}
\noindent We analyze elastic--pion nucleon scattering to third
order in the so--called small scale expansion. It is based on
an effective Lagrangian including pions, nucleons and deltas
as active degrees of freedom and counting external momenta, the pion mass
and the nucleon--delta mass splitting as small parameters. The fermion
fields are considered as very heavy. We present results for phase
shifts, 
threshold parameters and the sigma term. We discuss the convergence of the approach.
A detailed comparison with results obtained in heavy baryon chiral
perturbation theory to third and fourth order is also given.
\end{abstract}

\vspace{1cm}

\centerline{Keywords:~{\it Pion-nucleon scattering, effective field theory, delta resonance}}

\vfill

\end{titlepage}

\section{Introduction and summary}
\label{sec:intro}
\def\theequation{\arabic{section}.\arabic{equation}}
\setcounter{equation}{0}

Pion--nucleon scattering is an important testing ground for our understanding
of the chiral dynamics of QuantumChromoDynamics (QCD). It has been investigated to
third and fourth order in the chiral expansion, leading e.g. to precise
predictions for the threshold parameters~\cite{FMS,FM}. This approach is based on an effective
field theory with the active degrees of freedom being the asymptotically observable
pion and nucleon fields. When going to higher energies, the usefulness of the chiral
expansion is limited by the appearance of the nucleon resonances, the most prominent
and important of these being the $\Delta (1232)$ with spin and isospin 3/2. Its
implications for hadronic and nuclear physics are well established. Consequently,
one would like to have a consistent and systematic framework to include this
important degree of freedom in baryon chiral perturbation theory, as first
stressed by Jenkins and Manohar~\cite{JMdel}
and only recently formalized by Hemmert, Holstein and Kambor~\cite{HHK}. Counting
the nucleon--delta mass splitting as an additional small parameter, one arrives
at the so--called small scale expansion (which differs from the chiral expansion
because the $N\Delta$ splitting does not vanish in the chiral limit). It has already
been established that most of the low--energy constants appearing in the effective chiral
pion--nucleon Lagrangian are saturated by the delta~\cite{bkmlec} and thus the
resummation of such terms underlying the small scale expansion (SSE) lets one expect a better
convergence as compared to the chiral expansion. In addition, the radius of convergence
is clearly enlarged when including the delta as an explicit degree of freedom.
This of course increases the complexity of the approach since the pertinent effective
Lagrangian contains more structures consistent with all symmetries. The purpose of this
paper is to analyze pion--nucleon scattering to third order in this framework.
In particular, we want to address two questions. First, it has to be demonstrated
that for a given energy range, the third order SSE calculation leads to improved results
as compared to the ones obtained in the third order chiral expansion. Second, a precise
description of the resonant phase, i.e. of the $P_{33}$ partial wave, should be obtained.
As we will demonstrate, the SSE will pass both these tests. Furthermore, our
investigation can be used to test the convergence of the small scale
expansion. Indeed, from many models and more phenomenological approaches it is
believed that the most important delta contributions stem from the Born graphs
with intermediate resonance fields. We will address this issue in what
follows. Also, from the technical point of view, we provide for the first time a
systematic evaluation of the many $1/m$ corrections (using the machinery spelled
out in Ref.\cite{HHK}). Finally, we wish to point out that a systematic
Lorentz invariant formulation as it is available for the pion--nucleon
effective field theory~\cite{BL} is not yet available, but could be built
based on the pioneering investigation in Ref.~\cite{ET}. The heavy fermion approach
used here suffers from the same deficiencies as the standard heavy baryon
chiral perturbation theory, these problems are, however, not relevant to the
main issue we are going to address, namely the extension to higher energies.

\medskip\noindent
The pertinent results of the present investigation can be summarized as follows:
\begin{enumerate}
\item[(i)] We have constructed the  one--loop amplitude for
elastic pion--nucleon scattering based on an effective field theory
including pions, nucleons and deltas to third order in the small scale
expansion, ${\cal O}(\ve^3)$, where $\ve$ collects all small parameters
(external momenta, the pion mass and the nucleon--delta mass splitting).
We have constructed the pertinent terms of the effective Lagrangian including
the $1/m$ corrections. The amplitude contains  altogether 14 low--energy constants
from the nucleon, the nucleon--delta and the delta sector (if we count the
leading $\pi N \Delta$ coupling constant $g_{\pi N\Delta}$ as a LEC).
\smallskip
\item[(ii)]
The values of the LECs can be determined by fitting to the two S-- and four P--wave
amplitudes for different sets of available pion--nucleon phase shifts in the
physical region at low energies. We have performed two types of fits. In the first one, we fit
to the Matsinos phase shifts in the range of 40 to
100~MeV pion momentum in the laboratory frame. This allows for a direct comparison
with the results based on the chiral expansion. We find that the third order SSE
results are clearly better than the ones of the third order chiral expansion and
only slightly worse than the ones obtained at fourth order in the pion--nucleon EFT.
Second, we have fitted to the Karlsruhe phase shifts for pion lab momenta between
40 and 200~MeV. This allows for a better study of the resonance region.
In both cases, most fitted LECs are of ``natural'' size.
\smallskip
\item[(iii)]
We have studied the convergence of the small scale 
expansion by comparing the best fits based on the first (leaving the
coupling constant $g_{\pi N \Delta}$ free), second and third
order representation of the scattering
amplitudes. The third order corrections are in general not large, but they
improve the description of most partial waves. This indicates convergence
of the small scale expansion for this process. As anticipated, the most important contributions
come from the tree (Born) graphs with intermediate delta states. This allows
to pin down the coupling constant $g_{\pi N \Delta}$ fairly precisely.  
\smallskip
\item[(iv)]We can  predict the phases at {\it lower} and at
{\it higher} energies, in particular the threshold parameters
(scattering lengths and effective ranges). The results are not very
different from the third and fourth order studies based on the
chiral expansion, but the description of the
scattering length and the energy dependence  in the
delta channel are clearly improved. While the convergence of the isovector
S--wave scattering length is satisfactory to third order in the small scale
expansion, for drawing a conclusion on the isoscalar S--wave scattering length
a fourth order calculation is mandatory.
\smallskip
\item[(iv)]We have considered the pion--nucleon sigma term. To third order in the
small scale expansion, it depends on the low energy constant $c_1$ and the
coupling $g_{\pi N\Delta}$. For the KA85 phases, we get a value consistent
with previous determinations. We have also performed fits using the sigma
term extracted from a family of sum rules as input and shown that this
procedure leads to more reliable predictions for some LECs. 
\end{enumerate}

\medskip
\noindent
The manuscript is organized as follows.  In section~2, we discuss
the effective Lagrangian underlying our calculation. It decomposes
into separate nucleon, delta--nucleon and delta contributions.
We spell out all
details pertinent to our calculation including also the $1/m$ corrections.
Many of these terms have so far not been available in the literature.
Section~3 contains the results for the pion--nucleon scattering
amplitudes $g^\pm, h^\pm$ to third order in the small scale expansion.
In particular, we discuss the interplay between the low--energy constants
from the various sectors and the role of the so--called off--shell parameters.
The fitting procedure together
with the results for the phase shifts and threshold parameters are
presented in section~4. In particular, we give a detailed comparison
with the results obtained in the chiral expansion and discuss the
issue of convergence.  We also discuss the sigma term and show how it
can be used to further constrain the fits.
The pertinent Feynman rules, explicit expressions for the various $1/m$ corrections,
for the tree and loop contributions to the scattering amplitude and
the analytical expressions for the various threshold parameters are given in the appendix.

\section{Effective Lagrangian}
\label{sec:Leff}
\setcounter{equation}{0}

In this section, we briefly discuss the effective Lagrangian underlying
our calculation. We borrow heavily from the work of Hemmert et al.~\cite{HHK}
and refer the interested reader to that paper to fill in 
the details omitted here. The explicit
inclusion of the delta into an effective pion--nucleon field theory is
motivated by the fact that in certain observables this resonance plays a
prominent role already at low energies. The reason for this is twofold.
First, the delta--nucleon mass splitting \footnote{We use
the same symbol for the delta resonance as well as the $\Delta N$ mass
splitting. This cannot lead to confusion since from the context it is
always obvious what is meant.} is small,
\be
\Delta \equiv m_\Delta -m_N = 294~{\rm MeV} \simeq 3F_\pi~,
\ee
with $F_\pi = 92.4\,$MeV the weak pion decay constant. In the chiral
limit of vanishing quark masses, neither $\Delta$ nor $F_\pi$ vanish.
Therefore, such an extended EFT does not have the same chiral limit
as QCD, as it is well--known since long~\cite{GaZe}. Second, the delta
couples very strongly to the $\pi N \gamma$ system, e.g. the strong
$\Delta N \gamma$ M1 transition plays a prominent role in charged pion
photoproduction. One can set up a consistent power counting by
using the well--known heavy baryon techniques~\cite{JM,BKKM} and by treating 
the mass splitting $\Delta$ as an additional small parameter besides 
the external momenta and quark (meson) masses. Therefore, any matrix element
or transition current has a low energy expansion of the form
\be
{\cal M} = \ve^{n} \, {\cal M}_1 + \ve^{n+1} \, {\cal M}_2 +
\ve^{n+2} \, {\cal M}_3 + {\cal O}(\ve^{n+3})~,
\ee
where the power $n$ depends on the process under consideration 
(for pion--nucleon scattering, $n$ equals one)
and $\ve$ collects the three different small parameters,
\be
\ve \in \left\{ {q\over \Lambda_\chi},  {M_\pi\over \Lambda_\chi},
 {\Delta \over \Lambda_\chi} \right\}~,
\ee
with $\Lambda_\chi \simeq 1\,$GeV the scale of chiral symmetry
breaking, $M_\pi$ the pion mass,
and $q$ some external momentum. This power
counting scheme is often called $\ve$-- or small scale expansion (SSE).
As it is common in heavy baryon approaches,
the expansion in the inverse of the heavy mass and with respect to the
chiral symmetry breaking scale are treated simultaneously (because $m_N \sim
m_\Delta \sim \Lambda_\chi$).  The effective Lagrangian has the following 
low--energy expansion
\be
{\cal L}_{\rm eff} = {\cal L}^{(1)} + {\cal L}^{(2)} + {\cal L}^{(3)} + \ldots
\ee
where each of the terms ${\cal L}^{(n)}$ decomposes into a pure nucleon
($\pi N$), a nucleon--delta ($\pi N \Delta$) and  a pure delta part
($\pi\Delta$),
\be
 {\cal L}^{(n)} =   {\cal L}^{(n)}_{\pi N} + {\cal L}^{(n)}_{\pi N\Delta} +
  {\cal L}^{(n)}_{\pi\Delta}~, \quad n= 1,2,3, \ldots~.
\ee
We will now discuss these in succession. We note that the $N\Delta$ terms have to
be understood symmetrically, i.e. we can have an in--going nucleon and
an out--going delta or the other way around. The coupling of external fields
is done by standard methods, for our purpose we only have to consider a scalar
source to deal with the explicit chiral symmetry breaking due to the quark masses.

\subsection{Single nucleon sector}

This part is fairly standard, for completeness we collect the terms
necessary for the following discussion.  To lowest (first) order, the
relativistic pion--nucleon Lagrangian takes the form
\be
{\cal L}_{\pi N}^{(1)} = \bar{\Psi}_N ( i \not{\!\!D} - m +\frac{g_A}{2}  
\not{u} \gamma_5 ) \Psi_N~,
\ee
where the bi--spinor $\Psi_N$ collects the proton and neutron fields,
$g_A = 1.26$ is the axial--vector coupling constant and $m=m_N$ the
nucleon mass. All parameters in the Lagrangian should be taken at 
their chiral limit values. We do not exhibit this by special symbols
but it should be kept in mind. The heavy baryon projection is
most economically done using path integral methods as outlined in
Ref.\cite{BKKM}. This leads to a set of matrices, which organize the
transitions between the light--light ($N-N$), light--heavy ($N-h$) and
heavy--heavy ($h-h$) 
components of the nucleon fields, denoted by ${\cal A}$, ${\cal B}$ and ${\cal C}$, in
order,
\beq
{\cal L}_{\pi N} =  \bar{N} {\cal A}_N N + \bar{h} {\cal B}_N N 
  + \bar{N} \gamma_0 {\cal B}^\dagger_N \gamma_0 h - \bar{h} {\cal C}_N h~.
\eeq
The inverse of ${\cal C}$ is then further expanded in inverse powers of the
nucleon mass, leading to the decoupling of the positive and negative
velocity sectors. These matrices have a chiral expansion, i.e.
${\cal A} = {\cal A}^{(1)} +  {\cal A}^{(2)} + \ldots\,$. From the lowest order, we only need  
\bea
{\cal C}_{N}^{(0)} & = & 2 m ~, \\
{\cal A}_{N}^{(1)} & = & i v\cdot D + g_A S\cdot u ~, \\
{\cal B}_{N}^{(1)} & = & - \gamma_5 (2 i S\cdot D +\frac{g_A}{2} v\cdot u )~, \\
{\cal C}_{N}^{(1)} & = & i v\cdot D + g_A S\cdot u ~,
\eea
in terms of the nucleon four--velocity $v_\mu$ and the Pauli--Lubanski
spin--vector $S_\mu$ (for more details, see e.g. the review~\cite{BKMrev}).
We now turn to the second order terms. In the isospin limit of equal 
up and down quark masses, one has to deal with 4 operators,
\be
{\cal L}_{\pi N}^{(2)} = \bar{\Psi}_N  \Bigg[
c_1 \lanl \chi_+ \ranl 
-\frac{c_2}{8 m^2} \Big( \lanl u_\mu u_\nu \ranl \{D_\mu, D_\nu\} + {\rm h.c.} \Big)
+\frac{c_3}{2} \lanl u^2 \ranl +\frac{i c_4}{4} \sigma_{\mu\nu} [u_\mu,u_\nu] \Bigg]
\Psi_N~,
\ee
where $\chi_+$ includes the explicit chiral symmetry breaking and
traces in flavor space are denoted by $\langle \ldots \rangle$.
More precisely, $\chi_\pm = u^\dagger \chi u^\dagger \pm u
\chi^\dagger u$, $u_\mu = i(u^\dagger \partial_\mu u - u \partial_\mu
u^\dagger)$ and $U(x) =u^2(x)$ collects the pion fields (for details, see
e.g.~\cite{BKMrev}).
Note, however, that due to the presence of the delta degrees of
freedom the numerical values of these LECs are different from
the ones obtained in the pure pion--nucleon EFT. This is discussed in
more detail in Ref.\cite{BFHM}. From the second order transition
matrices we need the following terms:
\bea
{\cal A}_{N}^{(2)} & = & c_1 \lanl \chi_+ \ranl +c_2  (v\cdot u)^2
+c_3 u^2 
+c_4 [S^\mu,S^\nu] u_\mu u_\nu ~, \\
{\cal B}_{N}^{(2)} & = & - c_4 \gamma_5 [v\cdot u, S\cdot u] ~,\\
{\cal C}_{N}^{(2)} & = & -\Bigg[ c_1 \lanl \chi_+ \ranl +c_2  (v\cdot u)^2
+c_3 u^2 
+c_4 [S^\mu,S^\nu] u_\mu u_\nu \Bigg] ~.
\eea
Finally, from the third order Lagrangian, we only need one $1/m$ correction 
to the dimension two operator $\sim c_2$ and the following dimension
three operators 
\bea
{\cal A}_N^{(3)} & = &  \Big( i \,\,\frac{c_2}{2 m} \lanl (v\cdot u) u_\mu \ranl \, D^\mu + {\rm h.c.} \Big)
+ i d_1  [u_\mu,[v\cdot D,u^\mu]]  
+ i d_2  [u_\mu,[D^\mu,v\cdot u]]  \nonumber \\
&&+ i d_3  [v\cdot u,[v\cdot D,v\cdot u]]
+ d_5  [\chi_-,v\cdot u ] 
- i\,[S^\mu,S^\nu] d_{14} \langle [v\cdot D,u_\mu] u_\nu \rangle \nonumber \\
&&- i\,[S^\mu,S^\nu] d_{15}  \langle u_\mu [D_\nu,v\cdot u] \rangle 
+d_{16} \langle \chi_+ \rangle S\cdot u
+i\,d_{18} [S\cdot D,\chi_-]~.
\eea
Note that in $\pi N$ scattering, some of the LECs only appear in
certain combinations, here  $d_1 + d_2$
and $d_{14}-d_{15}$ are of relevance. Some of these LECs are needed
for the renormalization and their finite parts depend on the
regularization scale. We do not further specify this but it should be
kept in mind. Furthermore, there are some additional LECs just needed for
the renormalization, as spelled out e.g. in Ref.\cite{FMS}. We refrain from
writing down such terms here. As stressed before, the
numerical values of the finite parts of these LECs are influenced by
the presence of the delta and thus can not be taken over from the pure $\pi
N$ EFT. The remaining $1/m$ corrections from the elimination of the
small components are standard and can be found in Ref.\cite{FMS}.

\subsection{N$\Delta$--sector}

We now turn to the sector of the nucleons coupled to deltas and pions
as well as external sources. Here, we only consider external scalar
sources related to the explicit chiral symmetry breaking.
The first order relativistic Lagrangian
consistent with the requirement of point transformation invariance
takes the form
\be
{\cal L}_{\pi N\Delta}^{(1)} = g_{\pi N \Delta} \Bigg[ 
\bar{\Psi}_\mu^i \Theta_{\mu\alpha}(z_0) w_\alpha^i \Psi_N
+ \bar{\Psi}_N w_\alpha^i \Theta_{\alpha\mu}(z_0) \Psi_\mu^i\Bigg]~,
\ee
with $ \Theta_{\alpha\mu}(z_0) = g_{\mu\nu} + z_0 \gamma_\mu \gamma_\nu$,
${\Psi}_\mu^i$ is a conventional Rarita--Schwinger spinor and 
$w_\alpha^i = \frac{1}{2} \langle \tau^i u_\alpha\rangle$.
We have to deal with two parameters, the leading pion--nucleon--delta
coupling constant $g_{\pi N \Delta}$ and the so--called off--shell
parameter $z_0$. The latter will be discussed in more detail below.
The heavy baryon projection is more tedious since the delta
field has a large and a small spin--3/2 as well as four (off--shell)
spin--1/2 components. To keep only the large (light) spin--3/2 part,
one has to insert appropriate spin--isospin projection operators.
The technology to do that is spelled out in Ref.\cite{HHK}.
The effective Lagrangian has the genuine form
\beq
{\cal L}_{\pi N \Delta}  =  \bar{T} {\cal A}_{N \Delta} N 
  + \bar{G} {\cal B}_{N \Delta} N 
  + \bar{T} \gamma_0 {\cal D}^\dagger_{N \Delta} \gamma_0 h 
  + \bar{G} \gamma_0 {\cal C}_{N \Delta}^\dagger \gamma_0 h +{\rm h.c.}~,
\eeq
i.e. one has to deal with four types of transition matrices. These
are denoted by ${\cal A}_{N\Delta}$ (light--nucleon ($N$) to light--delta transitions ($T$)),
${\cal B}_{N\Delta}$ (light--nucleon ($N$) to heavy--delta transitions ($G$)),
${\cal C}_{N\Delta}$ (heavy--delta transitions ($G$) to heavy--nucleon ($h$)), and
${\cal D}_{N\Delta}$ (light--delta transitions ($T$) to heavy--nucleon
($h$)). Note that  the field $G$ has 5 components. 
To order $\ve$, these transition matrices read
\bea
{\cal A}_{N\Delta}^{(1)} & = & P_+ g_{\pi N \Delta} \ _3P_{\mu\alpha} 
w_\alpha^i P_+ ~, \\
{\cal B}_{N\Delta}^{(1)} & = & g_{\pi N \Delta} \left(
\begin{array}{c}
0~\\
-\frac{4(1+3 z_0)}{3} P_+ S_\mu  S\cdot w^i \,P_+ \\
2 z_0 P_- \gamma_5 v_\mu S\cdot w^i \,P_+ \\
-2 z_0 P_- \gamma_5 S_\mu v\cdot w^i \,P_+ \\
(1+z_0) P_+ v_\mu v\cdot w^i \,P_+ 
\end{array}
\right)~,\\
{\cal D}_{N\Delta}^{(1)} & = & 0~,\\
{\cal C}_{N\Delta}^{T\,(1)} & = & g_{\pi N \Delta} \left(
\begin{array}{c}
P_- \,w_\alpha^i \,\ _3P_{\alpha\mu} P_-\\
2 z_0 P_- v\cdot w^i S_\mu \gamma_5 \,P_+\\
(1+z_0)  P_- v\cdot w^i v_\mu P_-\\
\frac{-4(1+3 z_0)}{3} P_- S\cdot w^i S_\mu P_-\\
- 2 z_0 P_- S\cdot w^i v_\mu \gamma_5 \,P_+
\end{array}
\right)~,
\eea
with 
\be
\ _3P_{\mu\nu} = g_{\mu\nu} - v_\mu v_\nu -\frac{4}{1-d} S_\mu S_\nu ~,
\ee
in $d$ space--time dimensions and the $P_\pm$ are the usual velocity projection operators.
The second order relativistic $\pi N \Delta$
Lagrangian has two terms of relevance to our study, these read
\be
{\cal L}_{\pi N\Delta}^{(2)} = \bar{\Psi}_\mu^i \Theta_{\mu\alpha}(z) \Bigg[ 
i b_3  w_{\alpha\beta}^i \gamma^\beta
+i \frac{b_8}{m}  w_{\alpha\beta}^i i D^\beta \Bigg] \Psi_N + {\rm h.c.}~,
\ee
with $z$ another off--shell parameter also discussed below. To be more
precise, any new structure which appears with a low energy constant
has a separate off--shell parameter. However, these can be absorbed
in the corresponding LECs (as discussed in more detail below) and
we thus collectively call these new off--shell parameters $z$. The LECs
$b_i$ ($i=3,8$) are finite and will appear in the tree contribution
to the $\pi N$ scattering amplitude.  The heavy baryon projection is
done in terms of the appropriate second order transition matrices
\bea
{\cal A}_{N\Delta}^{(2)} & = & P_+ \ _3P_{\mu\alpha} 
i (b_3+b_8) w_{\alpha\beta}^{i} v_\beta  ~, \\
{\cal B}_{N\Delta,1}^{(2)}
& = & -P_- \ _3P_{\mu\alpha} \gamma_5 
2 i b_3  w_{\alpha\beta}^{i} S_\beta
P_+ ~, \\
{\cal }B_{N\Delta,2}^{(2)} & = & -\frac{4}{3} P_+ S_\mu \Big[
(1+3 z) i (b_3+b_8) w_{\alpha\beta}^{i}S^\alpha v^\beta 
-3 z  i b_3 w_{\alpha\beta}^i v_\alpha S_\beta\Big] ~, \\
{\cal B}_{N\Delta,3}^{(2)} & = & P_-v_\mu \gamma_5 \Big[
-(1+z) 
2 i b_3 w_{\alpha\beta}^{i}v^\alpha S^\beta
+2 z 
i (b_3 + b_8) w_{\alpha\beta}^i S_\alpha v_\beta\Big]~, \\
{\cal B}_{N\Delta,4}^{(2)} & = & \frac{4}{3}P_- S_\mu  \gamma_5 \Big[
(1+3 z) 
2 i b_3 w_{\alpha\beta}^i S_\alpha S_\beta
-\frac{3}{2} z 
i(b_3+b_8) w_{\alpha\beta}^i v_\alpha v_\beta \Big]~, \\
{\cal B}_{N\Delta,5}^{(2)} & = & P_+ v_\mu \Big[
(1+z) 
i(b_3 + b_8) w_{\alpha\beta}^i v_\alpha v_\beta 
-2 z 
i b_3 w_{\alpha\beta}^i S_\alpha S_\beta\Big] ~, \\
{\cal D}_{N\Delta}^{(2)} & = & 
P_- \gamma_5  
2 i b_3 S_\beta w_{\beta\alpha}^i 
 \ _3 P_{\alpha\mu} P_+~,\\
{\cal C}_{N\Delta}^{(3)} & = & 
P_+ \ _3P_{\mu\alpha} \frac{i b_8}{m} w_{\alpha\beta}^i i D_\beta~.
\eea
Note that at this order the first non--vanishing contribution to
${\cal D}_{N\Delta}$ appears. Finally, at third order, we only need the
relativistic effective Lagrangian and the corresponding transition
matrix ${\cal A}_{N\Delta}^{(3)}$. These read:
\bea
{\cal L}_{\pi N\Delta}^{(3)}& = \bar{\Psi}_{\mu'}^i \Theta_{\mu'\mu}(z) &\Bigg[ 
\frac{f_1}{m} [D_\mu,w_{\alpha\beta}^i] \gamma_\alpha i D_\beta 
-\frac{f_2}{2 m^2} [D_\mu,w_{\alpha\beta}^i] \{D_\alpha,D_\beta\}
\no\\&&
+f_4 w_\mu^i \lanl \chi_+ \ranl
+f_5 [D_\mu ,i \chi_-^i ] \Bigg] 
\Psi_N + {\rm h.c.}~,
\eea
\bea
{\cal A}_{N\Delta}^{(3)} & = & i{b_8\over m} w_{\mu\nu}^i i D_\nu +
(f_1+f_2) [D_\mu,w_{\alpha\beta}^i ] v_\alpha v_\beta
+f_4 w_\mu^i \lanl \chi_+ \ranl
+f_5 [D_\mu, i\chi_-^i ]  ~.
\eea
Four new LECs, which we call $f_i$,
appear. However, only the combinations $f_1 + f_2$ and $2f_4 -f_5$ are
of relevance in case of pion--nucleon scattering. Altogether,
in the $N\Delta$ sector we have 5 LECs  and the same number of
(unobservable) off--shell parameters. We have not counted $b_3$ and
$b_8$ separately, since
the latter can be absorbed in other LECs as detailed below.
The Feynman rules for the resulting $\pi N \Delta$ vertex
to first, second and third order are collected in appendix~\ref{app:vertex}.
The $1/m$ corrections originating from the elimination of the various
``small'' components are discussed below.

\subsection{Single $\Delta$ sector}
As in the previous paragraphs, we start with the dimension one effective
Lagrangian coupling the massive spin-3/2 fields to  pions,
\bea
{\cal L}_{\pi\Delta}^{(1)} &= -\bar{\Psi}_\mu^i &
[ (i \not{\!\!D}^{ij} - m_\Delta \delta^{ij}) g_{\mu\nu}
-\frac{1}{4} \gamma_\mu \gamma^\lambda (i \not{\!\!D}^{ij} - m_\Delta \delta^{ij})
\gamma_\lambda \gamma_\nu \no\\&&
+\frac{g_1}{2} g_{\mu\nu} \not{u}^{ij} \gamma_5 
+\frac{g_2}{2} (\gamma_\mu u_\nu^{ij} + u_\mu^{ij} \gamma_\nu )\gamma_5 
+\frac{g_3}{2} \gamma_\mu \not{u}^{ij} \gamma_5 \gamma_\nu ] 
\Psi_\nu^j~,
\eea
with $u_\mu^{ij} = u_\mu \, \delta^{ij}$ and $D_\mu^{ij}$ the appropriate
chiral covariant derivative. If one now performs simultaneously the $1/m$
expansion of the nucleon and delta fields, one can only rotate away the
nucleon mass, so that the $N\Delta$ mass splitting remains in the delta
propagator. Since this is, however, a quantity of order $\ve$, it can be
expanded systematically. In the heavy fermion approach, the Lagrangian
takes the form
\beq
{\cal L}_{\pi \Delta}  =  \bar{T} {\cal A}_\Delta T + \bar{G} {\cal B}_\Delta T 
  + \bar{T} \gamma_0 {\cal B}^\dagger_\Delta \gamma_0 G - \bar{G} {\cal C}_\Delta G~.
\eeq
The corresponding contributions from the three
types of transition matrices ${\cal A}_{\Delta}$, ${\cal B}_{\Delta}$, and
${\cal C}_{\Delta}$ are given by
\bea
{\cal A}_{\Delta}^{(1)} & = & -(i v\cdot D^{ij}-\Delta \delta^{ij} 
+ g_1 S\cdot u^{ij} ) g_{\mu\nu}~,  \\
{\cal B}_{\Delta,1}^{(1)} & = & P_- \ _3P_{\mu\nu} \, (2 i S\cdot D^{ij} 
                         +\frac{g_1}{2} v\cdot u^{ij} ) \gamma_5 \, P_+~, ~ 
{\cal B}_{\Delta,2}^{(1)}  =  - P_+ \, (\frac{2}{3} g_1 + g_2 ) S_\mu u_\nu^{ij} \, P_+ \no\\
{\cal B}_{\Delta,3}^{(1)} & = & P_- \, \frac{g_2}{2} v_\mu u_\nu^{ij} \gamma_5 \, P_+~, ~
{\cal B}_{\Delta,4}^{(1)}  =   \frac{4}{3} P_- S_\mu i D_\nu^{ij} \gamma_5 \, P_+~, ~
{\cal B}_{\Delta,5}^{(1)}  =  0~, \\
{\cal C}_{\Delta,11}^{-1\,(0)} & = & -\frac{1}{2m}  P_- \ _3P_{\mu\nu}  P_- ~, ~
{\cal C}_{\Delta,22}^{-1\,(0)} = \frac{1}{2m}  P_+ \ _1P_{\mu\nu} P_+  ~, ~
{\cal C}_{\Delta,23}^{-1\,(0)}  = - \frac{1}{2m} \frac{2}{3} P_+ S_\mu v_\nu
\gamma_5 P_-~, \no \\
{\cal C}_{\Delta,32}^{-1\,(0)} & = &  \frac{2}{3} P_- v_\mu S_\nu \gamma_5 P_+  ~, ~
{\cal C}_{\Delta,33}^{-1\,(0)} =  - \frac{1}{2m} P_- \ _2P_{\mu\nu} P_-  ~, ~
{\cal C}_{\Delta,44}^{-1\,(0)} =  - \frac{1}{2m} P_- \ _1P_{\mu\nu} P_- ~,\\
{\cal C}_{\Delta,45}^{-1\,(0)} & = & -\frac{1}{2m} 2 P_- S_\mu v_\nu \gamma_5 P_+ ~, ~
{\cal C}_{\Delta,45}^{-1\,(0)} = \frac{1}{2m}  2 P_+ v_\mu S_\nu \gamma_5 P_-  ~, ~
{\cal C}_{\Delta,55}^{-1\,(0)}  =  -\frac{1}{2m} \frac{5}{3} P_+ \ _2P_{\mu\nu}
P_+~,
\no \\
{\cal C}_{\Delta,11}^{-1\,(1)}  & = & \left(\frac{1}{2m}\right)^2 \ P_-{}\ _3P_{\mu \alpha}\ 
   (i v\cdot D^{i j} + \Delta \delta^{i j} + g_1 S\cdot u^{i j}) \ {}_3P_{\alpha \nu} \ P_-~,\no\\
{\cal C}_{\Delta,12}^{-1\,(1)}  & = & \left(\frac{1}{2m}\right)^2 \-2 P_- \ {}_3P_{\mu \alpha} 
   \ i S\cdot D^{i j} \gamma_5\  {}_1 P_{\alpha \nu} \ P_+~,\no\\
{\cal C}_{\Delta,13}^{-1\,(1)} & = & -\left(\frac{1}{2m}\right)^2 \ \frac{2}{3} \ P_- 
   \ {}_3P_{\mu \alpha} \ i D_\alpha^{i j}\  v_\nu \ P_-~,\no\\
{\cal C}_{\Delta,14}^{-1\,(1)} & = & \left(\frac{1}{2m}\right)^2 \ \frac{2 g_1}{3} P_- 
   \ {}_3P_{\mu \alpha} u_\alpha^{i j} S_\nu P_-~,\no\\
{\cal C}_{\Delta,15}^{-1\,(1)} & = & -\left(\frac{1}{2m}\right)^2 \ (g_1 + \frac{2}{3} g_2) 
   P_- \ {}_3 P_{\mu \alpha} u_{\alpha}^{i j} \gamma_5 v_\nu P_+~,\no\\
{\cal C}_{\Delta,22}^{-1\,(1)} & = & \left(\frac{1}{2m}\right)^2 \ \frac{8}{3} P_+ 
   S_\mu [ - i v\cdot D^{i j} + 2 \Delta \delta^{i j} 
   -\frac{1}{3} S\cdot u^{i j} (g_1 - 4 g_2 + 8 g_3) ] S_\nu P_+~,\no\\
{\cal C}_{\Delta,23}^{-1\,(1)} & = & \left(\frac{1}{2m}\right)^2 \ \frac{4}{3} P_+ 
   S_\mu [i v\cdot D^{i j} +\frac{1}{3} S\cdot u^{i j} (g_1 + 2 g_2) ] \gamma_5 v_\nu P_-~,\no\\
{\cal C}_{\Delta,24}^{-1\,(1)} & = & \left(\frac{1}{2m}\right)^2 \ P_+ S_\mu 
   [ - \frac{16}{9} i S\cdot D^{i j} - \frac{4}{3} v\cdot u^{i j} ( g_1 + 2 g_2) ] 
   \gamma_5 S_\nu P_-~,\no\\
{\cal C}_{\Delta,25}^{-1\,(1)} & = & \left(\frac{1}{2m}\right)^2 \ \frac{1}{9} P_+ 
   S_\mu [ 40 i S\cdot D^{i j} +v\cdot u^{i j} (14 g_1 + 16 g_2 + 16  g_3) ] v_\nu P_+~,\no\\
{\cal C}_{\Delta,33}^{-1\,(1)} & = & \left(\frac{1}{2m}\right)^2 \ P_- v_\mu 
   [ \frac{2}{3} i v\cdot D^{i j} + \frac{4}{3} \Delta \delta^{i j} +  
   \frac{10}{9} g_1 S\cdot u^{i j}  ] v_\nu P_-~,\no\\
{\cal C}_{\Delta,34}^{-1\,(1)} & = & \left(\frac{1}{2m}\right)^2 \ P_- v_\mu 
   [ -\frac{40}{9} i S\cdot D^{i j} -\frac{2}{3} g_1 v\cdot u^{i j}] S_\nu P_-~,\no\\
{\cal C}_{\Delta,35}^{-1\,(1)} & = & \left(\frac{1}{2m}\right)^2 \ P_- v_\mu 
   [ 4 i S\cdot D^{i j} +\frac{1}{3} v\cdot u^{i j} (g_1 -2 g_2)] \gamma_5 v_\nu P_+~,\no\\
{\cal C}_{\Delta,44}^{-1\,(1)} & = & \left(\frac{1}{2m}\right)^2 \ \frac{8}{3} P_- 
   S_\mu [ i v\cdot D^{i j} - 2 \Delta \delta^{i j} +\frac{5}{3} g_1 S\cdot u^{i j}] 
   S_\nu P_-~,\no\\
{\cal C}_{\Delta,45}^{-1\,(1)} & = & \left(\frac{1}{2m}\right)^2 \ P_- S_\mu 
   [ -\frac{4}{3} i v\cdot D^{i j} +\frac{16}{3} \Delta \delta^{i j} + 
   S\cdot u^{i j} (-4 g_1 + \frac{8}{9} g_2) ] \gamma_5 v_\nu P_+~,\no\\
{\cal C}_{\Delta,55}^{-1\,(1)} & = & \left(\frac{1}{2m}\right)^2 \ P_+ v_\mu 
   [ - \frac{2}{3}  i v\cdot D^{i j} + 4 \Delta \delta^{i j} + 
   \frac{2}{9} S\cdot u^{i j} (-17 g_1 + 12 g_2 - 8 g_3) ] v_\nu P_+~.
\eea
The other elements of ${\cal C}_\Delta^{-1}$ are given by
\begin{equation}
{\cal C}^{-1}_{\Delta,j i} = \gamma_0 {\cal C}^{-1\,\dagger}_{\Delta,i j} \gamma_0~.
\end{equation}
From the second order relativistic Lagrangian, we only need one term,
which is the mass insertion on the $\Delta$ propagator, i.e. the
analog to the dimension two nucleon term $\sim c_1$, 
\beq
{\cal L}_{\pi\Delta}^{(2)}  = \bar{\Psi}_{\mu}^i \Theta_{\mu\mu'}(z_1)
\, a_1 \lanl \chi_+ \ranl \delta^{ij}
\, g_{\mu'\nu'} \Theta_{\nu'\nu}(z_1')\Psi_{\nu'}^j \no\\
\eeq
which translates into
\bea
{\cal A}_{\Delta}^{(2)} & = & P_+ \ _3 P_{\mu\mu'} \,
a_1 \lanl \chi_+ \ranl  \, g_{\mu'\nu'} \delta^{ij} \ _3 P_{\nu'\nu} P_+~, 
\\
{\cal B}_{\Delta}^{(2)} & = & 0~.
\eea
For our purpose, no dimension three operator of the $\pi \Delta$
sector is needed, thus
\beq
{\cal A}_{\Delta}^{(3)}  = 0~.
\eeq
The genuine $1/m$ corrections will be considered in the next section.

\subsection{1/m corrections}

To calculate the $1/m$ corrections related to the elimination of the ``small''
components, it is most economical to use the path integral formalism
outlined in Ref.\cite{BKKM}. Here, an algebraic complication due to the
multiplication of degrees of freedom to be eliminated and also due to the
transition between the nucleon and the delta sectors appears. The machinery to do that
is spelled out in great detail in Ref.\cite{HHK}, here we only collect the
pertinent results and basic definitions. We note however that we give the most
detailed list of such terms so far worked out, only a subset of these is found
in  Ref.\cite{HHK}. After the canonical change of variables to make the
action quadratic in the fields, the various pieces of the effective Lagrangian
take the form:
\bea\label{shif1}
\tilde{{\cal L}}_{\pi N} & = & \bar{N} {\cal A}_N N
+\bar{N} \Big[ \gamma_0 \tilde{\cal B}_N^\dagger \gamma_0 \tilde{\cal C}_N^{-1} 
  \tilde{\cal B}_N + \gamma_0 {\cal B}_{ N \Delta}^\dagger\gamma_0 {\cal C}_\Delta^{-1} 
  {\cal B}_{N \Delta} \Big] N~,\\
\tilde{{\cal L}}_{\pi \Delta} & = & \bar{T} {\cal A}_\Delta T
+\bar{T} \Big[ \gamma_0 {\cal B}_\Delta^\dagger \gamma_0 {\cal
C}_\Delta^{-1} 
  {\cal B}_\Delta + \gamma_0 \tilde{\cal D}_{N \Delta}^\dagger\gamma_0
\tilde{\cal C}_N^{-1}
  \tilde{\cal D}_{N \Delta} \Big] T~,\\
\tilde{{\cal L}}_{\pi N \Delta} & = & \bar{T} {\cal A}_{N \Delta} N
+\bar{T} \Big[ \gamma_0 \tilde{\cal D}_{N \Delta}^\dagger \gamma_0 \tilde{\cal C}_N^{-1} 
  \tilde{\cal B}_N + \gamma_0 {\cal B}_{\Delta}^\dagger \gamma_0{\cal C}_\Delta^{-1} 
  {\cal B}_{N \Delta} \Big] N +{\rm h.c.}~,
\eea
employing the definitions:
\bea\label{shift}
\tilde{\cal B}_N& = & {\cal B}_N + {\cal C}_{N \Delta} {\cal C}_\Delta^{-1} {\cal B}_{N \Delta}~,\\
\tilde{\cal C}_N& = & {\cal C}_N - {\cal C}_{N \Delta} {\cal C}_\Delta^{-1}
{\cal C}_{N \Delta}^\dagger ~,\\
\tilde{\cal D}_{N \Delta}& = & {\cal D}_{N \Delta} +
 {\cal C}_{N \Delta} {\cal C}_\Delta^{-1} {\cal B}_{\Delta}~.
\eea
We note that the only $1/m$ corrections for ${\cal L}_{\pi \Delta}$ we need are 
propagator insertions. From these equations our previously made remark that
the low energy constants of the single nucleon sector get modified due to the
presence of the delta becomes quite obvious since the light nucleon to light
nucleon Lagrangian is modified by the appearance of the nucleon to
delta transition operators, e.g. the last term in Eq.(\ref{shif1}).
We come back to this point below. The algebra to work out these
terms is somewhat tedious and we collect the final results in appendix~\ref{app:1/m}.

\section{Pion--nucleon scattering}
\setcounter{equation}{0}

\subsection{Basic definitions}

In this section, we only give a few basic definitions pertinent to
elastic pion--nucleon scattering. For a more detailed discussion, we
refer to Ref.\cite{FMS}. In the center-of-mass system (cms), the amplitude for the process
$\pi^a(q_1) + N(p_1) \to \pi^b(q_2) + N(p_2)$ takes the
following form (in the isospin basis, with $a,b$ denoting cartesian
isospin indices of the pions): 
\beqa 
T^{ba}_{\pi N} &=& 
\biggl(\frac{E+m}{2m}\biggr) \, \biggl\lbrace \delta^{ba} 
\Big[ g^+(\omega,t)+ i \vec
\sigma \cdot(\vec{q}_2\times \vec{q}_1\,) \, h^+(\omega,t) \Big]
\nonumber \\ && \qquad\quad
+i \, \epsilon^{bac}
\tau^c \Big[ g^-(\omega,t)+ i \vec \sigma \cdot(\vec{q}_2 \times \vec{q}_1\,) \,
h^-(\omega,t) \Big] \biggr\rbrace 
\eeqa
with $\omega = v\cdot q_1 = v\cdot q_2$ the pion cms energy, 
$E_1 = E_2 \equiv E = ( \vec{q\,}^2 +m^2)^{1/2}$ the nucleon energy and
$\vec{q\,}_1^2 = \vec{q\,}_2^2 \equiv \vec{q\,}^2 = [(s-M_\pi^2-m^2)^2 -4m^2M_\pi^2]/ (4s)$.
$t=(q_1-q_2)^2$ is the invariant momentum transfer squared and $s$ 
denotes the total cms energy squared. 
Furthermore, $g^\pm(\omega,t)$ refers to the
isoscalar/isovector non-spin-flip amplitude and $h^\pm(\omega,t)$ to the
isoscalar/isovector spin-flip amplitude. This form is most suitable
for a heavy fermion calculation (as done here) since it is already defined in a
two--component framework.

\medskip\noindent
The quantities of interest are the partial wave amplitudes
$f_{l\pm}^\pm (s)$, where $l$ refers to the
orbital angular momentum, the superscript '$\pm$' to the isospin (even
or odd)
and the subscript '$\pm$' to the total angular momentum ($j=l\pm s$), are given
in terms of the invariant amplitudes via
\beq
f_{l\pm}^\pm (s) = {E+m \over 16 \pi \sqrt{s}} \, \int_{-1}^{+1} dz \, \biggl[\,
g^\pm \, P_l (z) + \vec{q\,}^2 \, h^\pm \, (P_{l\pm1}(z) -zP_l (z) ) \biggr]~,
\eeq
where $z = \cos(\theta)$ is related to the scattering angle.
The $P_l(z)$ are the conventional Legendre polynoms. For a given
isospin $I$, the phase shifts
$\delta_{l\pm}^I (s)$ can be extracted from the partial waves via
\beq
f_{l\pm}^I (s) = {1\over 2i | \vec{q\,}|} \, \biggl[ \exp(2i
\delta_{l\pm}^I (s))  - 1 \biggr]~.
\eeq
For vanishing inelasticity, which is the case for the energy range
considered in this work ($\sqrt{s} \le 1.3\,$GeV),  
the phase shifts are real. They are given
by 
\beq\label{unit}
\delta_{l\pm}^I (s) = \arctan \bigl( |\vec{q}\,|\, 
{\rm Re}~f_{l\pm}^I (s) \bigr)~.
\eeq
In the low energy region, one could equally well use the definition without the arctan, the difference
being of higher order. For the phase shifts in the kinematical region considered
here, this difference is not negligible. In fact, this arctan
prescription is nothing but a unitarization procedure which is
mandated by the appearance of the poles in the delta propagator at
$\omega = \Delta$. Although there is some arbitrariness in this
unitarization procedure, it has already been shown in
Ref.\cite{BKMres} that it leads to the proper resonance width (for
sharp resonances like the delta or the $\rho$ in pion--pion
scattering, for more details see Ref.\cite{BKMres}). 
Consequently, the phase shifts presented in what follows are
based on Eq.(\ref{unit}).

\subsection{Small scale expansion of the amplitudes}

\begin{figure}[H]
\centerline{
\epsfysize=2.7in
\epsffile{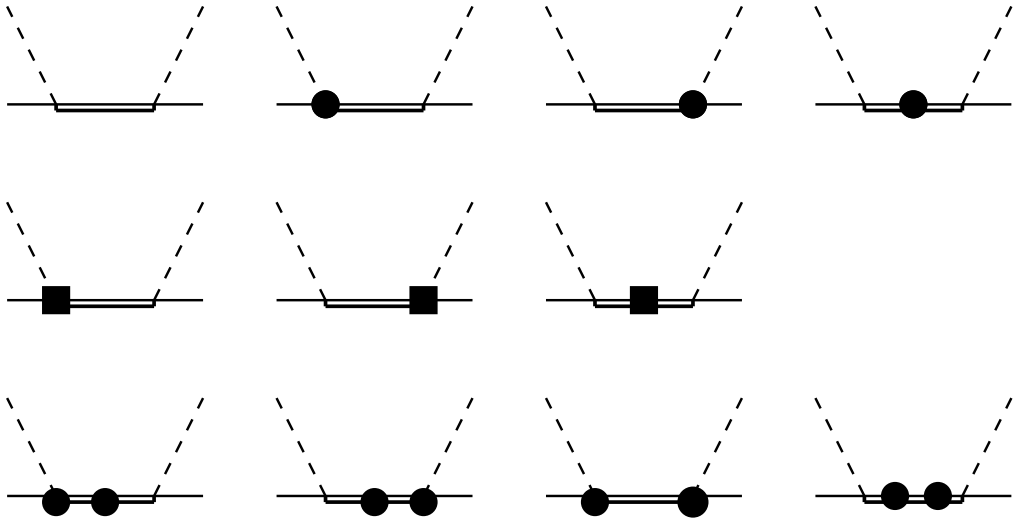}
}
\vskip 2.5cm
\caption{
Tree and counterterm graphs involving the delta. Dashed, solid and double lines
refer to pions, nucleons and deltas, in order. Crossed graphs and
diagrams that vanish are not shown. The vertex dot and vertex square
refer to an insertion from the dimension two, respectively three effective
$\pi \Delta$ or $\pi N\Delta$ Lagrangian.\label{fig:tree}  
}

\end{figure}

\begin{figure}[H]
\centerline{
\epsfysize=7in
\epsffile{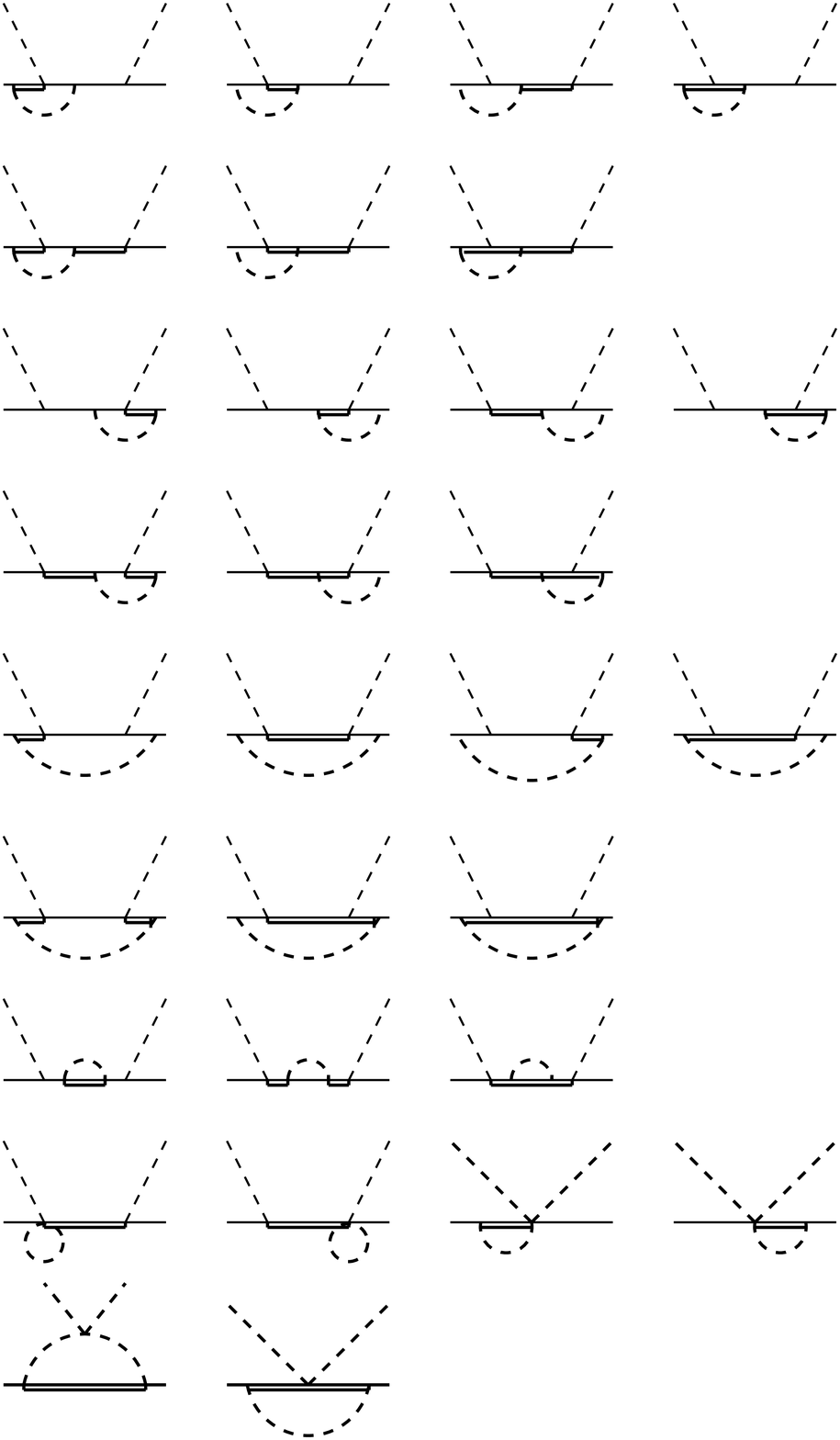}
}
\vskip 1.5cm
\caption{
Leading one loop graphs involving the delta. Dashed, solid and double lines
refer to pions, nucleons and deltas, in order. Crossed graphs and
diagrams that vanish are not shown.\label{fig:loop}  
}

\end{figure}

In this section we discuss  the small scale expansion of the non-spin-flip
and spin-flip amplitudes
$g^\pm, h^\pm$. These consist of essentially three pieces, which are
the tree contributions with intermediate nucleons and deltas, counterterm parts of polynomial type
as well as the unitarity corrections due to the pion loops (again with intermediate
nucleons and deltas). The tree plus counterterm and loop graphs with intermediate deltas are shown
in Fig.\ref{fig:tree} and Fig.\ref{fig:loop}, respectively. We remark that the $\pi\pi N\Delta$
vertices
only start at order ${\cal O}(\ve^2)$ and thus all loop graphs of relevance here 
have $N\Delta$ couplings
linear in the pion field. Formally, the small scale expansion of the various
amplitudes has the form
\beq\label{chexp}
X = X^{\rm tree} + X^{\rm ct} + X^{\rm loop} \,\, , \quad
X= g^\pm  ,  h^\pm \,\, ,
\eeq
where the tree contribution subsumes all Born terms with fixed coefficients,
and the counterterm amplitude the ones
proportional to the dimension two and three  LECs.
The last term in Eq.(\ref{chexp}) is the leading one--loop
amplitude consisting of terms of order $\ve^3$.
The latter is a complex--valued function and
restores unitarity in the perturbative sense. Its various terms
are all proportional to $1/F_\pi^4$.
Note that we treat the chiral symmetry breaking scale $\Lambda_\chi \simeq
1\,$GeV on the same footing as the nucleon and delta mass.
These amplitudes are functions of two kinematical variables,
which we choose to be the pion energy and the invariant
momentum transfer squared, i.e. $X= X (\omega, t)$. In what
follows, we mostly suppress these arguments. It is also instructive
to compare these amplitudes with the ones of the pure pion--nucleon EFT,
based on the {\it chiral expansion}. These terms are, of course, contained in the
SSE to third order and their explicit expressions are given in Ref.\cite{FMS}
(we remind the reader that the numerical values of the LECs are different, for
the reasons discussed above).\footnote{In this paper, we only display the novel delta
contributions to the amplitudes, threshold parameters and so on. The purely nucleonic
terms can be found in  Ref.\cite{FMS}.}
The novel tree and loop terms, shown in Fig.\ref{fig:tree}
and Fig.\ref{fig:loop}, are of fourth order (or higher) in the chiral expansion, since the
delta is frozen out in that approach and generates second (and higher) order contact
interactions (as discussed in more detail in Ref.\cite{bkmlec}). Therefore, the amplitudes
based on the third order small scale expansion can be considered as partial fourth order chiral
calculations. Of course, they also contain higher order terms in the chiral expansion,
since the delta is not frozen out. One can therefore expect that in most channels the
third order SSE calculation should lead to a better description of the phase shifts than
given by the third order chiral expansion. Such an expectation is indeed borne out by the
actual results to be presented below.

\medskip\noindent
The full one--loop amplitude to order $\ve^3$ is obtained after mass
and coupling constant renormalization,
\beq
(\krig{g}_A , \krig{m}, \krig{\Delta}, F , M)
\to ({g}_A , {m} , \Delta ,  F_\pi , M_\pi )~.
\eeq
The pion mass, decay constant and Z--factor are standard,
\bea
M_\pi^2 & = & M^2 \Bigg\{ 1 +\frac{2 M^2}{F^2} \ell_3 +\frac{\Delta_\pi}{2 F^2}
\Bigg\}~,
\\
Z_\pi & = & 1-\frac{2 M^2}{F^2} \ell_4 - \frac{\Delta_\pi}{F^2}~,
\\
F_\pi & = & F \Bigg\{ 1 +\frac{M^2}{F^2} \ell_4 -\frac{\Delta_\pi}{F^2} \Bigg\}~,
\eea
in terms of the LECs $\ell_{3,4}$ and the divergent pion self--energy tadpole
$\Delta_\pi$ given e.g. in Ref.\cite{BKMrev}.
Similarly, for the nucleon mass shift, the nucleon Z--factor and the
pion--nucleon coupling, we use (for more details, see Refs.\cite{FM,BFHM}):
\bea
m & = & \krig{m} -4 M^2 c_1 +\frac{9}{4} \left( \frac{g_A}{F} \right)^2 J_2(0)\no\\&&
+ \frac{4}{3} \left( \frac{g_{\pi N \Delta}}{F} \right)^2 
\left[ (M^2-\Delta^2) \left( J_0(-\Delta) -\frac{2 \Delta}{M^2} \Delta_\pi \right)
+ \frac{\Delta}{48 \pi^2} (3 M^2-2 \Delta^2)  \right]~,
\\ 
Z_N & = & 1- 8 M^2 \tilde{d}_{28} (\lambda)
+\frac{9}{4} \left( \frac{g_A}{F} \right)^2 J_2'(0)
-\frac{3 M^2}{32 \pi^2} \left( \frac{g_A}{F} \right)^2 \no\\&&
+ \frac{4}{3} \left( \frac{g_{\pi N \Delta}}{F} \right)^2 
\left[ (M^2-\Delta^2)  J_0'(-\Delta) +2 \Delta J_0(-\Delta)
\right.\no\\&&\qquad\qquad\qquad\left.
+\frac{2}{M^2}(M^2-3 \Delta^2) \Delta_\pi
+\frac{1}{16\pi^2} (-M^2+2 \Delta^2)\right]  ~,
\\
\frac{g_A}{F_\pi} & = & \frac{\krig{g}_A}{F} \Bigg\{
1-\frac{M^2}{F^2} \ell_4 +\frac{4 M^2}{g_A} d_{16} (\lambda)
+\frac{g_A^2}{4 F^2} \left( \Delta_\pi -\frac{M^2}{4 \pi^2} \right)\no\\&&
+\left( \frac{g_{\pi N \Delta}}{F} \right)^2 \Bigg(
\frac{1}{3} \left( 4 -\frac{100}{81}\frac{g_1}{g_A} \right) 
  \left[ (M^2-\Delta^2) J'_0(-\Delta) +2 \Delta J_0(-\Delta)\right]\no\\&&
-\frac{32}{27 \Delta} \left[ M^2 J_0(0) -(M^2-\Delta^2) J_0(-\Delta)  \right]
+\frac{8}{27 M^2} \Delta_\pi \left[(M^2-19 \Delta^2) -\frac{25}{9} \frac{g_1}{g_A} (M^2-3 \Delta^2)\right]\no\\&&
+\frac{1}{24 \pi^2} (-M^2+2 \Delta^2) (2-\frac{70}{27}\frac{g_1}{g_A})
+\frac{2}{81 \pi^2} (-3 M^2+2 \Delta^2) \Bigg)
\Bigg\}~,
\eea
in terms of the standard loop functions $J_{0,2} (\omega)$~\cite{BKMrev}.
Finally, for the delta, we only need the mass shift,
\beq
m_\Delta  = \krig{m}_\Delta  -4 M^2 a_1 
-\frac{25}{108} \left( \frac{g_1}{F} \right)^2 M^2 J_0(0) 
-\frac{1}{3} \left( \frac{g_{\pi N \Delta}}{F} \right)^2 (M^2-\Delta^2)
\left( J_0(\Delta)+\frac{2 \Delta}{M^2} \Delta_\pi \right) \, .
\eeq
We refrain from giving the much more complicated expressions for the delta
Z--factor and the renormalization of $g_{\pi N \Delta}$.
Also, we always work with renormalized LECs, all infinities appearing in the
loop diagrams are  accounted for by the corresponding infinite parts of the
pertinent counterterms. What is still missing in the SSE is a systematic 
investigation of renormalization as it is the case in (heavy) baryon chiral
perturbation theory.

\subsection{Counterterm combinations and off--shell parameters}
\label{sec:cts}

First, we enumerate the novel low energy constants  related to the
$\pi \Delta$ and $\pi N \Delta$ sectors. Consider first the   $\pi N
\Delta$ couplings. At leading order, there is only $g_{\pi N \Delta}
=1.05$ from the width of the decay $\Delta \to
N\pi$~\cite{HHK}. Since the necessary resummation at the pole of
the delta includes some higher order terms, we will also perform fits
leaving this coupling free. We expect, however, that the so determined
value is not very different from 1.05. In addition, from the dimension
two and three Lagrangians, we have the LEC combinations $b_3+b_8$,
$f_1+f_2$ and $2f_4 - f_5$. We note that there is a $1/m$ correction to
$b_8$, nonetheless the LECs $b_3$ and $b_8$ need not be treated
separately because the term $b_8/ m$ can be absorbed in the dimension three LECs from the
nucleon sector, as detailed below. From the $\Delta\pi$ sector, we
only have the coupling $g_1$, which we leave free. In the large $N_c$
limit of QCD, one obtains the relation $g_1 = 9g_A/5$, with $g_A=1.26$  the axial coupling
constant measured in neutron $\beta$--decay. However, we will not use
this relation but rather consider the  value determined from the fit
to the data as a check of
the large $N_c$ expansion of QCD. In addition, there is the
dimension two LEC $a_1$. It only leads to a mass shift of the delta and
can thus be absorbed completely in the physical nucleon--delta mass
splitting. All these couplings are accompanied
by separate off--shell parameters, which we (collectively) have
denoted by $z_0$, $z$ and $z'$.\footnote{Note that the special
  treatment of the leading $\pi N \Delta$ coupling is done for
  historical reasons.} These off--shell parameters are not observable, so they can be
chosen freely. According to the common practice, we set $z=z'
=z_0=\frac{1}{2}$, which is of course  a little bit arbitrary.
As stated before, due to the explicit appearance of the $\Delta$ in the
theory, there are new $1/m$ corrections to ${\cal L}_{\pi N}$, which
can be absorbed 
into the LECs $c_i$, $d_i$ of the nucleon sector. This also means that
a certain part of these LECs is explained by $\Delta$ properties (as
it is well known, see Ref.\cite{bkmlec}). In the following, we will
not absorb these pieces in the $c_i$ and $d_i$ LECs (with one
exception to be given below), because such a
redefinition also depends on the off--shell parameters $z,z'$ and
$z_0$. If one were to perform this redefinition, it would take the form
\begin{eqnarray}
c_2 & \longrightarrow & c_2 -2 \frac{\gpnd^2}{9m} (3+4 z_0^2) + 8 \de \frac{\gpnd^2}{9 m^2} (1 +2 z_0^2)~, \\
c_3 & \longrightarrow & c_3 +\frac{\gpnd^2}{9m} (1+8 z_0 + 12 z_0^2) -2 \de \frac{\gpnd^2}{9 m^2} (1+6 z_0 +8 z_0^2) ~,\\ 
c_4 & \longrightarrow & c_4 +\frac{\gpnd^2}{9m} (1+8 z_0 +12 z_0^2) -2 \de \frac{\gpnd^2}{9 m^2} (1+6 z_0 +8 z_0^2) ~,\\ 
(\bar{d}_1 + \bar{d}_2 )& \longrightarrow & (\bar{d}_1 + \bar{d}_2) +\frac{1}{4} \frac{\gpnd^2}{9 m^2} (5+4 z_0^2) -2 \frac{\gpnd (b_3 + b_8)}{9m} z_0 
           -\frac{1}{2} \frac{\gpnd b_8}{9m} (1 +4 z +12 z z_0)~,\no \\ && \\ 
\bar{d}_3 & \longrightarrow & \bar{d}_3 -3 \frac{\gpnd^2}{9m^2} +\frac{1}{4} \frac{\gpnd (b_3+b_8)}{9m} (16 +5 z -4 z z_0) 
          -\frac{1}{4} \frac{\gpnd b_8}{9m} (4 +5 z -20 z z_0)~, \no \\ &&\\ 
\bar{d}_5 & \longrightarrow & \bar{d}_5 +\frac{1}{2} \frac{\gpnd^2}{9 m^2} (2-z_0) 
     -\frac{1}{8}\left( \frac{\gpnd (b_3+b_8)}{9m} - \frac{\gpnd
         b_8}{9m}\right) (6+13 z +4 z z_0) ~,
       \no \\ && \\
(\bar{d}_{14}-\bar{d}_{15}) & \longrightarrow & (\bar{d}_{14}-\bar{d}_{15}) -\frac{\gpnd^2}{9 m^2} (1+4 z_0^2) -8 \frac{\gpnd (b_3+b_8)}{9m} z_0 
           -2 \frac{\gpnd b_8}{9m} (1+4 z +12 z z_0)~.\no \\ && 
\end{eqnarray}
In this case, the $c_i$ now also have contributions proportional to $\Delta/m$. 
Since $c_2$ and $c_4$ also appear in the third order amplitude, using these renormalized LECs would lead
to contributions of fourth order, which is beyond the accuracy of our
calculation. This 
we consider another reason not to make this redefinition. 
Similar problems related to the mixing of various chiral
orders arise also in the  fourth order analysis performed in the pure
pion--nucleon EFT, see Ref.\cite{FM}. 
However, in one case, we have to perform this redefinition. The contribution
proportional to $b_8$ has to be absorbed into $(\bar{d}_1 +\bar{d}_2)$,
$\bar{d}_3$, $\bar{d}_5$ and
$(\bar{d}_{14} -\bar{d}_{15})$. The resulting LECs and combinations
thereof are labeled by a subscript ``$\Delta$''. If that is not done,
one introduces an additional redundant fit parameter. So this is the only
redefinition we are  making. The other effects from $b_8$ can then be
absorbed into the  following combinations
\beq
{\mbox (f_1+f_2) - \frac{b_8}{2m}} \quad {\rm and} \quad
{\mbox (2 f_4 -f_5) - \frac{b_8}{4m}}~.
\eeq
So we end up with the nine LECs from ${\cal L}^{(2,3)}_{\pi N}$ plus
four LECs (or combinations thereof) from ${\cal L}^{(2,3)}_{\pi N
  \Delta}$ (counting the leading $\pi N \Delta$ coupling as a free
parameter, although we also perform fits with its value fixed)
and one LEC from the $\Delta \pi$ sector.

\section{Results}
\setcounter{equation}{0}

\subsection{The fitting procedure}
There are various possibilities to fix the LECs.
We proceed here along the similar lines as in Refs.\cite{FMS,FM}, namely we fit to the
phase shifts given by  different partial wave analyses in the low
energy region. This allows for a better comparison with the results obtained
in the chiral expansion. As input we use the phase shifts of the
Karlsruhe (KA85) group~\cite{koch} and from the analysis of
Matsinos~\cite{mats} (EM98). Of course, there is also the phase shift
analysis of the VPI/GW group, which we do not use here for two reasons.
First, the solution called SP98 from the VPI/GW
group~\cite{SAID} is no longer available (it was used in Refs.~\cite{FMS,FM})
and, second, no threshold parameters for the newest solution SP00 have been published.
Clearly, once these are available we can update our investigation, but we do not
believe that it will lead to wildly different results and new insight compared to what is
presented below. Since no uncertainties are available for the KA85 solution,
we mimic the ones of the Matsinos analysis in that case, which is 1.5\% for $S_{31}$,
0.5\% for $S_{11}$, 1\% for $P_{33}$ and 3.5\% for the other P--waves.
This assignment gives more weight to the better determined larger
partial waves and is more natural than one common global
error. We remark that the Matsinos analysis only includes data up to
200~MeV pion laboratory momentum. For this analysis, we proceed as in Refs.~\cite{FMS,FM},
namely fit to the S-- and P--wave phase shifts for momenta between 40 and 100~MeV
and predict the phases at lower and higher energies. For the KA85 solution, we extend
the fit region up to 200~MeV and thus predict the phases up to about 300~MeV (and of
course also in the threshold region). As in Refs.~\cite{FMS,FM} the
LEC  $\bar{d}_{18}$ is fixed by means of the Goldberger--Treiman discrepancy, i.e.
by the value for the pion--nucleon coupling constant extracted in the various
analyses. The actual values of $g_{\pi N}$ are $g_{\pi N} = 13.4\pm
0.1 \,$ and $13.18 \pm 0.12$
for KA85 and EM98, respectively.  For the terms including  the $\Delta$, we perform fits
with $g_{\pi N \Delta }$ fixed (at the value of 1.05) and letting it free (as discussed
above).  Throughout, we use $g_A = 1.26$, $F_\pi =
92.4\,$MeV, $m = 938.27\,$MeV, $\Delta = 294\,$MeV and $M_\pi = 139.57\,$MeV.

\subsection{Phase shifts and threshold parameters}

\renewcommand{\arraystretch}{1.1}
\begin{table}[hbt]
\begin{center}
\begin{tabular}{|c|c|c|c|c|}
    \hline
    LEC      &  Fit 1  &  Fit 1*  & Fit 2  & Fit 2*  \\
    \hline\hline
${c}_1$ & $ 0.77\pm 0.02 $ & $ 0.39\pm 0.02 $ & $-0.44\pm 0.01$ & $-0.32\pm 0.01$\\
${c}_2$ & $-17.9\pm 0.03 $ & $-15.5\pm 0.03 $ & $-0.67\pm 0.03$ & $-1.59\pm 0.03$\\
${c}_3$ & $ 20.2\pm 0.06 $ & $ 16.7\pm 0.04 $ & $-0.07\pm 0.03$ & $ 1.15\pm 0.03$\\
${c}_4$ & $-15.6\pm 0.04 $ & $-12.5\pm 0.03 $ & $-2.51\pm 0.04$ & $-3.44\pm 0.04$ \\
\hline
$(\bar{d}_1+\bar{d}_2)_\Delta$
      & $-5.91\pm 0.06 $ & $-5.81\pm 0.07 $ & $ 0.03\pm 0.04$ & $-0.36\pm 0.04$\\
$(\bar{d}_3)_\Delta$
      & $ 7.68\pm 0.06 $ & $ 6.60\pm 0.07 $ & $-0.93\pm 0.04$ & $-0.44\pm 0.04$\\
$(\bar{d}_5)_\Delta$
      & $-1.07\pm 0.03 $ & $-0.59\pm 0.04 $ & $ 0.75\pm 0.02$ & $ 0.71\pm 0.02$\\
$(\bar{d}_{14}-\bar{d}_{15})_\Delta$
      & $-5.18\pm 0.20 $ & $-0.09\pm 0.21 $ & $-0.44\pm 0.15$ & $-0.60\pm 0.15$\\
$\bar{d}_{18}$
      & $-0.98\pm 0.19 $ & $-0.97\pm 0.19 $ & $-1.35\pm 0.14$ & $-1.38\pm 0.14$\\
\hline
$g_{\pi N \Delta}$ & $ 1.32\pm 0.03$ & $ 1.05^\star  $ & $0.98 \pm 0.05$ & $ 1.05^\star  $\\
$b_3+b_8         $ & $-12.0\pm 0.10$ & $-11.1\pm 0.13$ & $0.51 \pm 0.09$ & $-0.28\pm 0.08$\\
$f_1+f_2-\frac{b_8}{2m}         $ & $ 29.3\pm 0.30$ & $ 32.4\pm 1.30$ & $-19.1\pm 0.57$ & $-18.6\pm 0.51$\\
$2f_4-f_5-\frac{b_8}{4m}        $ & $ 40.2\pm 0.73$ & $ 53.5\pm 0.92$ & $-30.3\pm 0.92$ & $-27.0\pm 0.82$\\
$g_1             $ & $-1.42\pm 0.02$ & $-2.65\pm 0.03$ & $-1.10\pm 0.04$ & $-0.94\pm 0.04$\\
\hline\hline
  \end{tabular}
  \caption{Values of the LECs in appropriate units of inverse GeV
           for the various fits described in the text.
           \label{tab:LEC}}
\end{center}\end{table}
\noindent
We are now in the position to present results.
For the analysis of Matsinos, we use 17 points for each
partial wave in the range of $q_\pi = 41.4 - 96.3\,$~MeV.
We have performed fits leaving the coupling $g_{\pi N \Delta}$ free
or fixing its value at 1.05. We call these fits~1 and 1*, respectively.
Therefore, we have to determine 14 and 13 parameters for these two fits.
Since the data basis of Ref.\cite{mats} includes only data up to 200 MeV,
we use these fits mostly for comparison with the results of the chiral
expansion presented in Refs.\cite{FMS,FM}. This is different for the
Karlsruhe data basis, which extends up to very high energies. Consequently,
for this case (KA85), we have fitted to the data up to 200~MeV pion  lab
momentum (i.e. 10 points per partial wave at $q_\pi = 40, 60, 79,
97, 112, 130, 153, 172, 185, 200$~MeV).  We  extend the fits to higher
energies than it was done in Refs.\cite{FMS,FM}; this is possible due
to the explicit inclusion of delta and allows to study explicitly the
resonance region. Again, we perform fits with varying and fixed $g_{\pi N \Delta}$,
which are denoted by fits~2 and 2*, respectively.
The resulting LECs are collected in table~\ref{tab:LEC}. We point out again that
the values for the LECs from the nucleon sector (the $c_i$ and $d_i$) can not
be compared with the ones obtained in the chiral expansion for the various
reasons discussed above.  For the Matsinos data, the $\chi^2$/dof is very small
and slightly better for fit~1 than for fit~1*. The reasons for this very small
 $\chi^2$/dof are discussed
in some detail in  Ref.\cite{FM}. For the KA85 phases, the $\chi^2$/dof is essentially
the same for both cases (and larger than for fits~1,1*, see again  Ref.\cite{FM}.).
We also note that while the value of $g_{\pi N \Delta}$
is somewhat larger than the canonical value of 1.05 for fit~1, it comes out slightly
lower in the case of fit~2. In the following, we will mostly discuss the results
of the fits~1*,2* with $g_{\pi N \Delta} = 1.05$. Most LECs come out
of natural size, i.e. of order one, with the exception of the LEC combinations
from ${\cal O}(\ve^3)$. This can be traced back to the fact that there are large
cancelations at this order between the loop graphs with intermediate deltas and the
counterterms. Such a phenomenon was also observed in the third order
SSE calculation of neutral pion photoproduction, see~\cite{BHM}. Also, in case
of fits~1,1*, the values for the dimension two LECs are quite large, and, in
particular, the value for $c_1$ is positive. Furthermore, it is important to stress that
we find some sizeable correlations between the LECs $c_{1,2,3}$. This
is not unexpected since the corresponding terms contribute only to the
small isoscalar S--wave scattering amplitudes (in certain combinations,
e.g. the S--wave isoscalar scattering length is only sensitive to
the LEC combination $-2c_1 + c_2 + c_3$). This will be taken up in the
next subsection. We note that the value for the $\pi\Delta$ coupling $g_1$
comes out very different from the large $N_c$ prediction $g_1 = 9g_A/5 = 2.27$.
Note, however, since $g_1$ only appears in the third order loop contribution,
one can not expect to pin it down very precisely.

\begin{figure}[H]
\centerline{
\epsfysize=7in
\epsffile{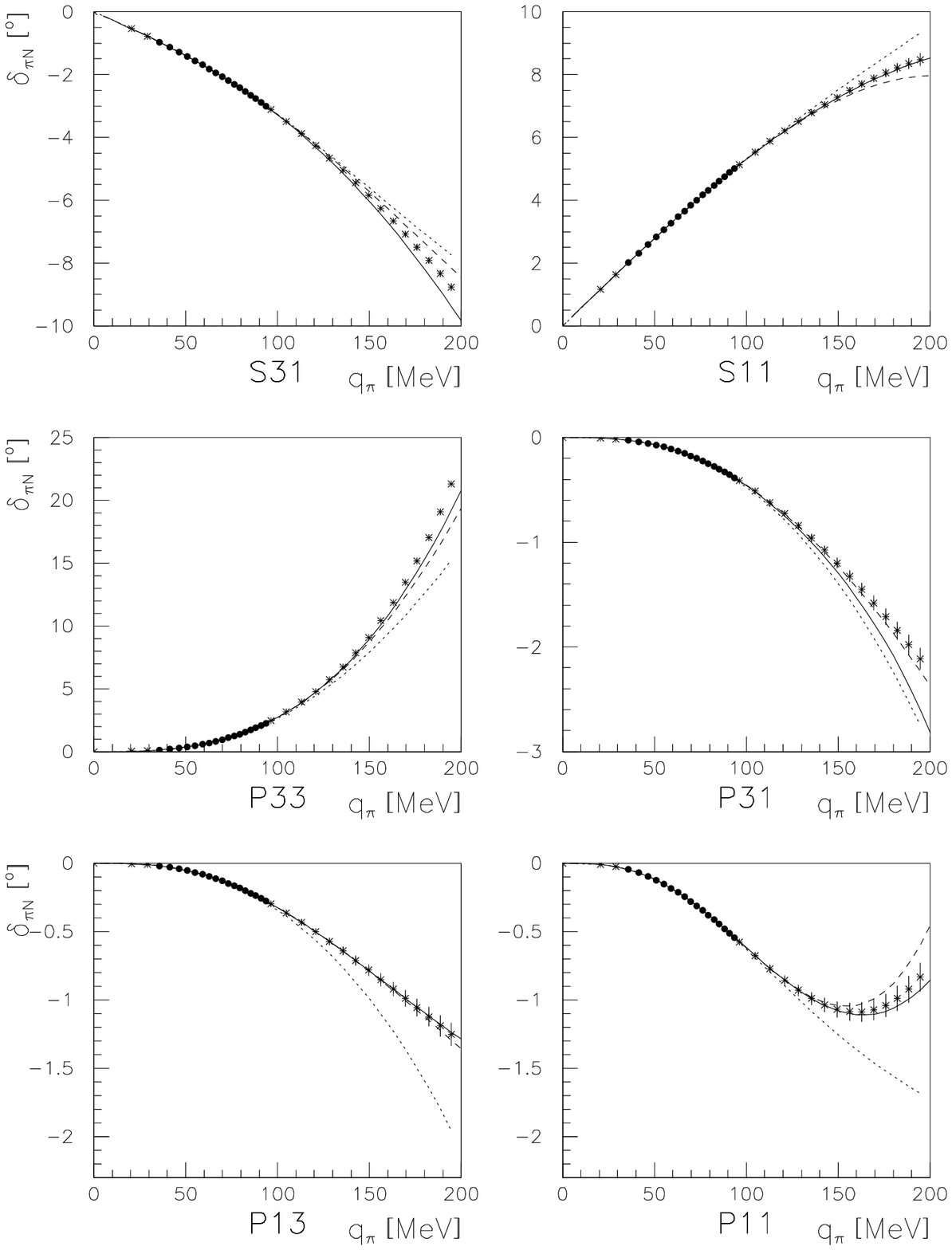}
}
\vskip 0.5cm
\caption{
Fits and predictions for the EM98 phase shifts as a function of 
the pion laboratory  momentum $q_\pi$.
Fitted in each partial wave are the data between 41 and 97~MeV (filled circles). For higher
and lower energies, the phases are predicted as shown by the solid lines.
Dotted and dashed lines: Third and fourth order calculation based on the
chiral expansion~\cite{FMS,FM}.}
\label{fig:EM98}
\end{figure}

\begin{figure}[bht]
\centerline{
\epsfysize=7in
\epsffile{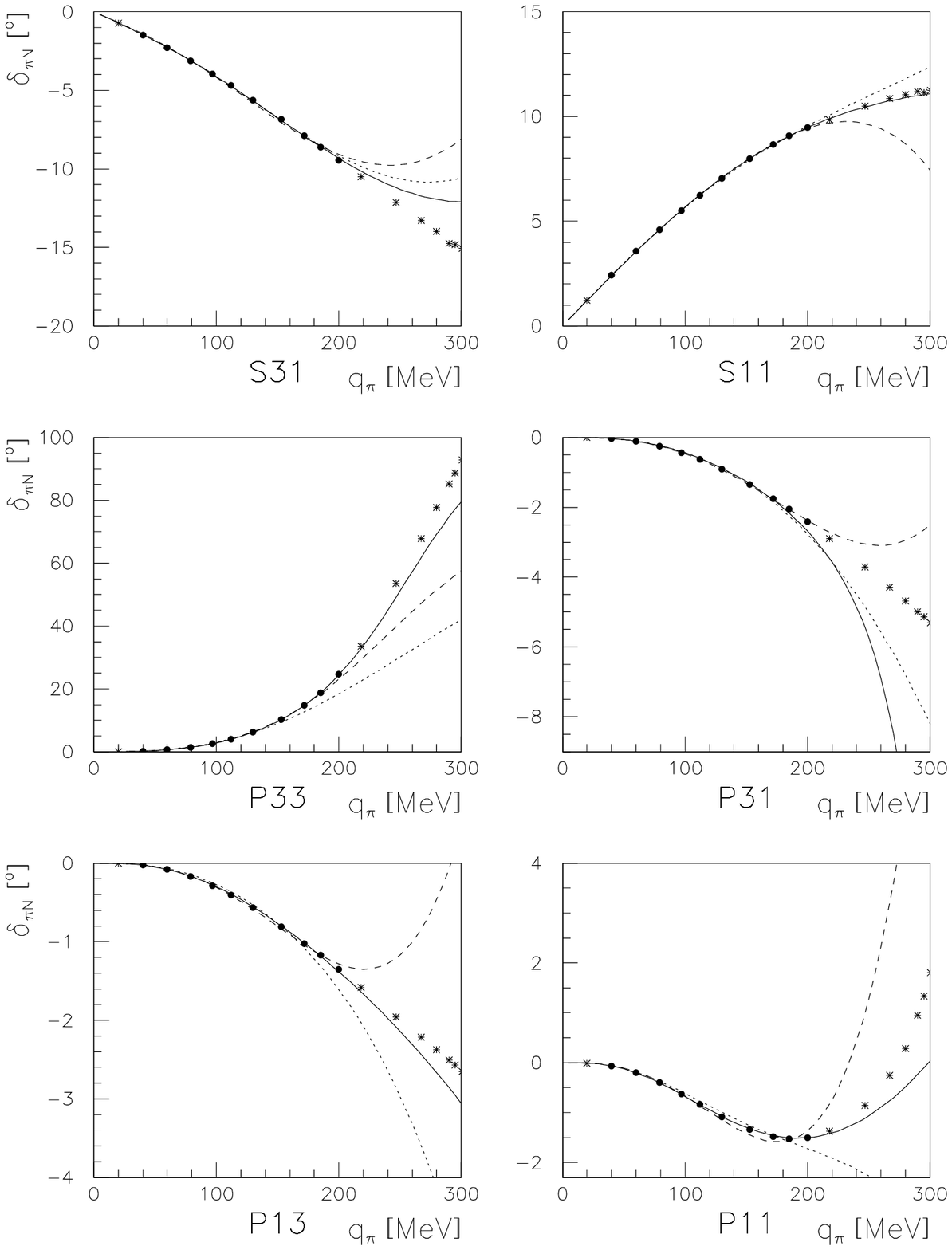}
}
\vskip 0.5cm
\caption{
Fits and predictions for the KA85 phase shifts as a
function of the pion laboratory  momentum $q_\pi$.
Fitted in each partial wave are the data between 40 and 200~MeV (filled
circles). For higher and lower energies, the phases are predicted as
shown by the solid lines.
Dotted and dashed lines: Third and fourth order calculation based on the
chiral expansion~\cite{FMS,FM}.
\label{fig:KA85}
}
\end{figure}

\medskip\noindent
The resulting fits and predictions based on the EM98 and KA85 phases are shown in Figs.\ref{fig:EM98}
and \ref{fig:KA85}, respectively, in comparison to the results based on the third and fourth order
chiral amplitudes. The number of LECs to be fitted was 9 and 14 in these cases. The results
of the third order SSE calculation are clearly better than the ones based on the third
order chiral expansion and comparable  to the fourth order results, although the overall
description is still slightly better in the latter case. As expected, the most prominent
improvement can be found in the $P_{33}$ partial wave, which is well described up to
the delta pole, see Fig.\ref{fig:KA85}. At lower energies, one can predict the threshold
parameters, which are collected in table~\ref{tab:thr}
for the fits~1* and 2*  in comparison with the direct
determination from the Matsinos and Karlsruhe phase shift analyses. The overall consistency
is satisfactory but not as good as in the case of the fourth order chiral expansion.
Again, we find a significant improvement of the scattering volume
in the $P_{33}$ channel, as expected due to the explicit inclusion of the delta.
\begin{table}[htb]
\begin{center}
\begin{tabular}{|c||c|c||c|c||}
    \hline
   Obs.    &  Fit 1    &  Fit 2  & EM98 & KA85  \\
    \hline\hline
$a^+_{0+}$  & $ 0.41$ & $-0.94$  & $ 0.41\pm 0.09 $  & $-0.83$ \\
$b^+_{0+}$  & $-4.22$ & $-4.60$  & $ -4.46  $        & $-4.40$ \\
$a^-_{0+}$  & $ 7.74$ & $ 8.95$  & $ 7.73\pm 0.06 $  & $ 9.17$ \\
$b^-_{0+}$  & $ 1.42$ & $ 1.63$  & $ 1.56   $        & $ 0.77$ \\
$a^+_{1-}$  & $-5.49$ & $-5.54$  & $-5.46\pm 0.10 $  & $-5.53$ \\
$a^+_{1+}$  & $13.08$ & $13.27$  & $ 13.13\pm 0.13 $ & $13.27$ \\
$a^-_{1-}$  & $-1.21$ & $-1.46$  & $-1.19\pm 0.08 $  & $-1.13$ \\
$a^-_{1+}$  & $-8.21$ & $-8.14$  & $-8.22\pm 0.07 $  & $-8.13$ \\
  \hline\hline
  \end{tabular}
  \caption{Values of the S-- and P--wave threshold parameters for the 
           fits~1* and 2* in comparison to the respective data.
           The results for fits~1 and 2 are similar and thus are not given.
           Units are appropriate inverse powers of the pion mass times 10$^{-2}$.
           \label{tab:thr}}
\end{center}
\end{table}

\begin{figure}[bht]
\centerline{
\epsfysize=7in
\epsffile{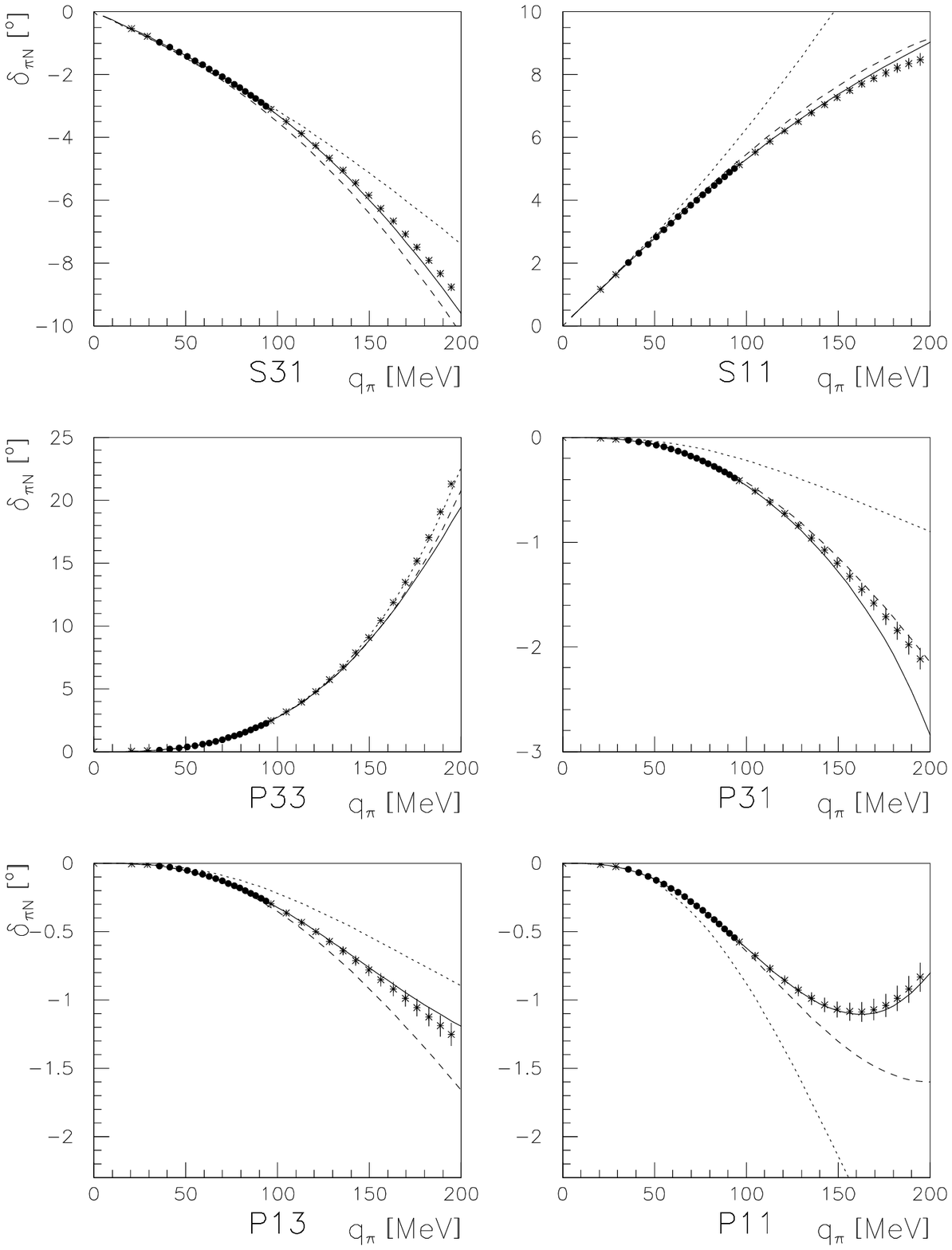}
}
\vskip 0.5cm
\caption{
Fits and predictions for the EM98 phase shifts  as a
function of the pion laboratory  momentum $q_\pi$ 
to first (dotted lines), second (dashed lines)
and third (solid lines) order in the small scale expansion.
Fitted in each partial wave are the data between 41 and 97~MeV (filled circles). For higher
and lower energies, the phases are predicted.
}
\label{fig:EMconv} 
\end{figure}

\begin{figure}[bht]
\centerline{
\epsfysize=7in
\epsffile{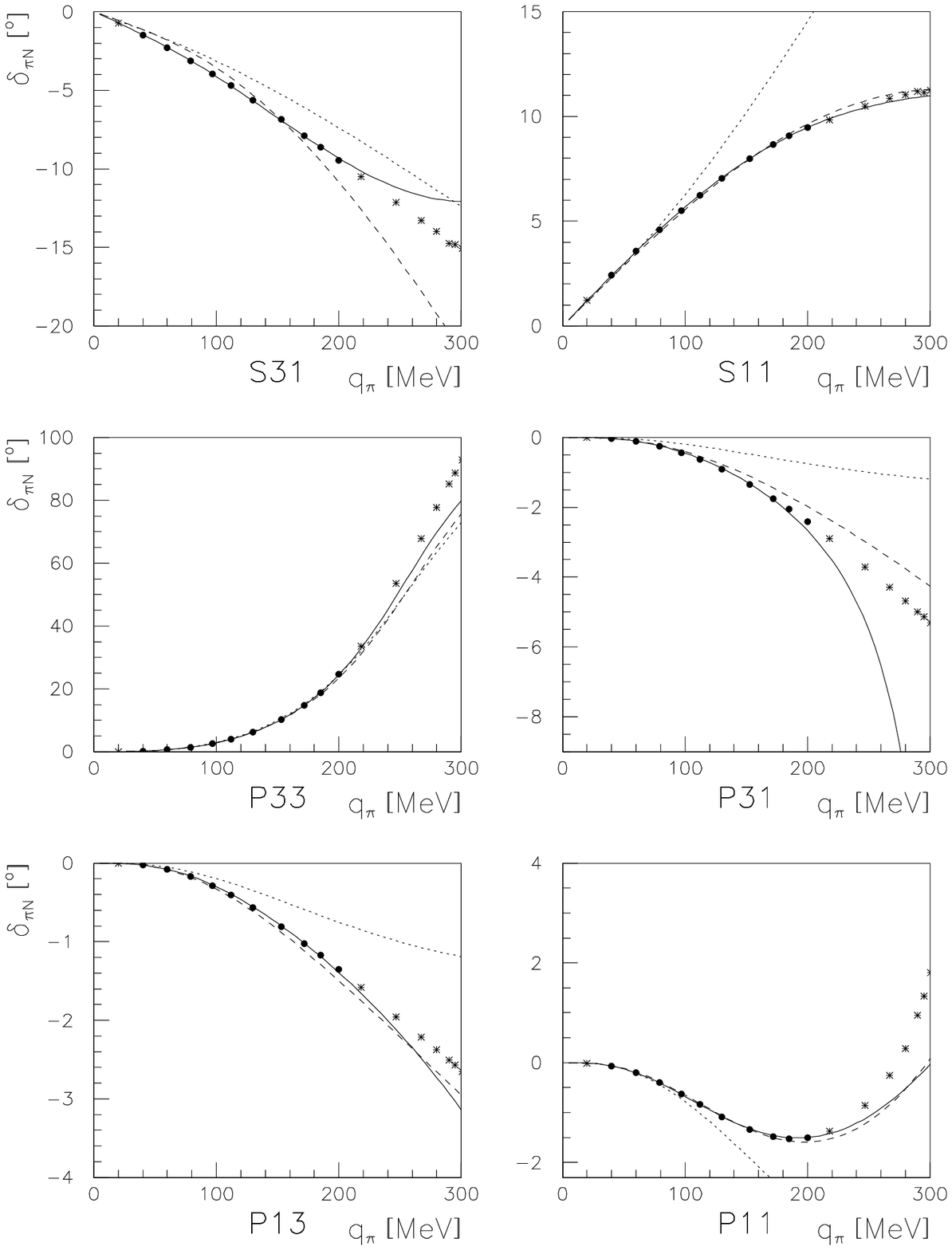}
}
\vskip 0.5cm
\caption{
Fits and predictions for the KA85 phase shifts as a
function of the pion laboratory  momentum $q_\pi$
to first (dotted lines), second (dashed lines)
and third (solid lines) order in the small scale expansion.
Fitted in each partial wave are the data between 40 and 200~MeV (filled circles).
For higher  and lower energies, the phases are predicted.
}
\label{fig:KAconv} 
\end{figure}

\medskip\noindent
It is also important to discuss the issue of convergence. For that, we redo the
fits for the amplitudes at first and second order in the small scale expansion.
To leading order, one only has the coupling $g_{\pi N \Delta}$, whereas at second
order one has the additional four LECs from the nucleon sector and one LEC combination
from ${\cal L}^{(2)}_{\pi N\Delta}$. The resulting phase shifts are shown in
Fig.\ref{fig:EMconv} for the EM98 analysis and in Fig.\ref{fig:KAconv} for the KA85
case. While the first order result is only good in the $P_{33}$ partial wave,
the second order fits are of comparable quality than the third order ones, although
at third order the  $\chi^2$/dof is significantly better. Such a behavior does not come completely
unexpected since one expects the delta Born graphs to play the most significant role.
Such an expectation is build on experience with many models that include the delta
or also the explicit calculations of Compton scattering off nucleons in the SSE~\cite{HHKK}.
This behavior is different from what is found in the chiral expansion, where one still has
large corrections when going from second to third order but mostly modest ones
from third to fourth order, see~\cite{FMS,FM}. This means that in the channels where
the delta plays a significant role, the resummation of higher order terms in the chiral
expansion is important and well described in the approach used here. It is also
interesting to study the convergence of the S--wave scattering lengths, as it
has been done for the chiral expansion in~\cite{FM}. In the small scale
expansion we obtain the results collected in table~\ref{tab:aconv1}. The
convergence for the isovector scattering length is similar to what is obtained
in the chiral expansion (as expected from the arguments presented in
Ref.\cite{bkmpin2}). The isoscalar S--wave scattering length receives a large
correction when going from second to third order, indicating that certain
fourth order pieces related to pion--nucleon physics are still missing (as can
also be inferred from the study in Ref.\cite{FM}). Some of these  results
were also found by Ellis and Tang~\cite{ET}, although their approach is based on a
relativistic treatment of the fermion fields.  Therefore, in their approach all
$1/m$ corrections are resummed, but the delta is treated in a less systematic manner.
Datta and Pakvasa~\cite{DP} had already used the one loop representation of Ref.\cite{moj}
and also added the delta, but not in a systematic fashion as done here. They also
found a much improved description of the  $P_{33}$ partial wave.
\begin{table}[htb]
\begin{center}
\begin{tabular}{|r|r|r|r|}
    \hline
          & ${\cal O}(\ve)$  &   ${\cal O}(\ve^2)$  &  ${\cal O}(\ve^3)$ \\
    \hline\hline
$a_{0+}^+ \quad$EM98 & 0.0  & 0.24  &    0.41  \\
                KA85 & 0.0  & 0.44  & $-$0.94  \\
\hline
$a_{0+}^- \quad$EM98 & 7.90 & 7.90  &    7.74  \\
                KA85 & 7.90 & 7.90  &    8.95  \\
\hline\hline
  \end{tabular}
  \caption{Convergence of the S--wave scattering lengths. 
   ${\cal O}(\ve^n)$ means that all terms up-to-and-including
   order $n$ are given. Units are 10$^{-2}/M_\pi$.
    \label{tab:aconv1}}
\end{center}\end{table}

\subsection{The sigma term}

So far, we have exclusively considered the amplitudes in the physical region. We have
noted that there are some strong correlations between some LECs.
To further address that problem, we need additional input. This is provided
e.g. by the so--called pion--nucleon sigma term,
which is the expectation value of the QCD symmetry breaking terms
in a proton (or neutron) state. It can be derived from the scalar
form factor of the nucleon,
\beq
\sigma (t) = \langle N(p') | \hat{m} (\bar u u + \bar d d)| N(p)
\rangle ~, \quad t = (p'-p)^2~,
\eeq
with $|N(p)\rangle$ a nucleon state of four--momentum $p$ and $\hat m$
the average light quark mass. The sigma term is nothing but the scalar
form factor at $t=0$, $\sigma \equiv \sigma(t=0)$. We remark that the
sigma term can not be obtained directly from scattering data. One
usually considers its value at the Cheng--Dashen point, $t=2M_\pi^2$,
where the chiral corrections are minimized. $\Sigma \equiv\sigma (t=2M_\pi^2)$ and
$\sigma$ differ by 17~MeV, with 15~MeV stemming from the scalar form 
factor~\cite{gls} and (at most) 2~MeV from the so--called
remainder~\cite{bkmcd}. In addition,  there exists a whole family of
relations between $\Sigma$ and certain combinations of threshold
parameters, as detailed in Ref.\cite{juerg}. These relations
have been worked out to third order in the chiral expansion.
We will use here the version given in Ref.\cite{glls},
\beq\label{sumG}
\Sigma = \pi F_\pi^2 [(4+2\mu+\mu^2)a_{0+}^+ -4M_\pi^2 b_{0+}^+
+ 12\mu M_\pi^2 a_{1+}^+ ] + \Sigma_0~,
\eeq
with $\Sigma_0 =-12.6\,$MeV and $\mu = M_\pi/m \simeq 1/7$. 
A special variant, which also contains some fourth order
pieces, has recently been given by Olsson~\cite{MO},
\beqa\label{sumO}
{\Sigma} &=& [{F_\pi^2}\, F(2M_\pi^2)]~, \\
 F(2M_\pi^2) &=& 14.5 \, a_{0+}^{+} - 5.06\, 
(a_{0+}^{1/2})^2 - 10.13\, (a_{0+}^{3/2})^2 - 16.65\, b_{0+}^{+}
- 0.06 \, a_{1-}^{+} + 5.70 \, a_{1+}^{+} - 0.05~,\nonumber
\eeqa
with the quantities on the right--hand--side being given in units 
of the pion mass. We will use these sum rules to further constrain
our LECs.

\medskip\noindent
To third order in the small scale expansion, the scalar from factor of
the nucleon reads (the scalar sector has also been discussed by Kambor~\cite{JK}):
\bea\label{sff}
\sigma(t) & = & -4 M^2 c_1 
- \left( \frac{g_A}{F} \right) ^2 \frac{3 M^2}{8}  \left[ -2 J_0(0) + (t-2 M^2) K_0(t,0)
\right]\no\\&&
+\left( \frac{g_{\pi N \Delta}}{F} \right)^2 \frac{2 M^2 }{3} 
\left[ 2 J_0(-\Delta) + (2 M^2 -t -2 \Delta^2) K_0(t,-\Delta) -2 \Delta I_0(t) 
+\frac{\Delta}{8 \pi^2} \right]~,\no\\&&
\eea
in terms of the standard loop functions listed in~\cite{moj}. For $t=0$, this gives
\bea\label{sigma}
\sigma(t=0) & = & -4 M^2 c_1  
-\left(\frac{g_A}{F} \right)^2\frac{9 M^3}{64 \pi}
+\left( \frac{g_{\pi N \Delta}}{F} \right)^2 \frac{M^2}{2 \pi^2} \sqrt{\Delta^2-M^2}
\ln{\left( \frac{\Delta}{M} + \frac{\sqrt{\Delta^2-M^2}}{M} \right)}~.\no\\&&
\eea
Note that the terms $\sim g_{\pi N \Delta}^2$ in
Eqs.(\ref{sff},\ref{sigma}) are of fourth order in the chiral
expansion, as already pointed out long time ago~\cite{bkmss}. We note
that $\sigma$ only depends on the LEC $c_1$ (and also the coupling 
$g_{\pi N \Delta}^2$ for the fits where it is left free). If we now
use $c_1$ as determined in fits~1 and 1*, we get an unphysical
negative sigma term because $c_1$ is positive and large. That,
however, is an artifact since not all the $c_i$'s can be determined 
independently. For the fits~2~(2*), we obtain the following values,
\beq\label{sigKA}
\sigma = 51.1~(47.3)~{\rm MeV}~.
\eeq
These numbers are slightly larger but consistent within error bars with
what has been found before, see Refs.\cite{gls,paul}. However, one
might question the accuracy of this determination because also in this
case one has large correlations between certain LECs. To overcome
this, we have performed a different set of fits. As additional input we 
take the sigma term as determined from the sum rules,
Eq.(\ref{sumG}) and Eq.(\ref{sumO}) using the
threshold parameters from the EM98 and the KA85 analysis as input. 
More precisely, since the two sum rules differ by some terms of fourth
order (and higher) in the chiral expansion, for each set of input
threshold parameters we get two numbers for $\sigma = \Sigma -
17\,$MeV, which we average to obtain the central value and their
spread is taken as the theoretical uncertainty. This gives 
$\sigma = (58.5 \pm 5.4)\,$MeV for EM98 and $\sigma = (45.5 \pm 2.7)\,$MeV
for KA85. The corresponding fits with this additional input are
denoted by fit~1$^\star$ and fit~2$^\star$.
\renewcommand{\arraystretch}{1.1}
\begin{table}[hbt]
\begin{center}
\begin{tabular}{|c|c|c|c|c|}
    \hline
    LEC      &  Fit 1$^\star$  &  Fit 2$^\star$   \\
    \hline\hline
${c}_1$ & $-0.18\pm 0.02 $ & $-0.35\pm 0.09 $ \\
${c}_2$ & $-5.72\pm 0.03 $ & $-1.49\pm 0.66 $ \\
${c}_3$ & $ 6.05\pm 0.03 $ & $ 0.93\pm 0.87 $ \\
${c}_4$ & $-8.93\pm 0.04 $ & $-3.08\pm 0.81 $ \\
\hline
$(\bar{d}_1+\bar{d}_2)_\Delta$
      & $ 5.52\pm 0.07 $ & $-0.57\pm 0.64 $ \\ 
$(\bar{d}_3)_\Delta$
      & $-4.12\pm 0.07 $ & $-0.30\pm 0.64 $ \\
$(\bar{d}_5)_\Delta$
      & $-0.89\pm 0.04 $ & $ 0.74\pm 0.10 $ \\
$(\bar{d}_{14}-\bar{d}_{15})_\Delta$
      & $-18.8\pm 0.22 $ & $-0.10\pm 1.60 $ \\
$\bar{d}_{18}$
      & $-0.99\pm 0.20 $ & $-1.34\pm 0.24 $ \\
\hline
$g_{\pi N \Delta}$ & $ 1.27\pm 0.04$ & $ 1.00\pm 0.08$ \\
$b_3+b_8         $ & $-7.33\pm 0.12$ & $-0.03\pm 0.89$ \\
$f_1+f_2-\frac{b_8}{2m}         $ & $-31.3\pm 0.98$ & $-17.9\pm 4.49$ \\
$2f_4-f_5-\frac{b_8}{4m}        $ & $-68.0\pm 0.77$ & $-23.0\pm 9.78$ \\
$g_1             $ & $-2.05\pm 0.02$ & $-1.05\pm 0.41$ \\
\hline\hline
  \end{tabular}
  \caption{Values of the LECs in appropriate units of inverse GeV
           for the fits using as additional input the sigma term.
           \label{tab:LECstar}}
\end{center}\end{table}
\noindent
The resulting LECs are shown in table~\ref{tab:LECstar}. First, as
expected from the numbers given in Eq.(\ref{sigKA}), there are only
minor changes in case of fit~2$^\star$ as compared to fits~2,2*. This
is very different for fit~1$^\star$ compared to fits~1,1*.
The value of $c_1$ is now negative and also, the pion--nucleon
dimension two LECs have more natural values. In addition, the
correlations between the LECs $c_{1,2,3}$ are somewhat smaller
than before. A very important result is the stability of 
the coupling constant $g_{\pi N\Delta}$, which comes out consistent
with what was found in fits~1 and 2, respectively.

\bigskip

\subsection*{Acknowledgments}
We are grateful to Thomas Hemmert for many useful comments and to Evgeny
Epelbaum for checks on some parts of the calculation. N.F. thanks all
members of the Kellogg Radiation Lab at Caltech for hospitality
extended to her during a stay when part of this work was performed.

\bigskip\bigskip
\appendix
\section{Pion--nucleon--delta vertices}
\def\theequation{\Alph{section}.\arabic{equation}}
\setcounter{equation}{0}
\label{app:vertex}

Here, we give the Feynman rules for the relevant $\pi N \Delta$
vertices. Consider first the case on an
incoming nucleon (momentum $p_1$), an outgoing delta (momentum $P_\mu$, isospin $i$), 
and an outgoing pion (momentum $q$, isospin $a$):\\
\underline{1st order:}
\be
\frac{g_{\pi N \Delta}}{F} q^\mu \delta^{ia}~.
\ee
\underline{2nd order:}
\be
{\bf -}\frac{b_3+b_8}{F} v\cdot q q^\mu \delta^{ia}
-\frac{g_{\pi N \Delta}}{m F} P_\mu v\cdot q \delta^{ia}~.
\ee
\underline{3rd order:}
\bea
\delta^{ia} \frac{1}{F} & \Bigg\{
&- q_\mu \Big( (e_1+e_2) (v\cdot q)^2 +2 M^2 (-2 e_4 +e_5) \Big)
\no\\&&
+\frac{q_\mu}{m} \Big[
\frac{b_3}{2} (v\cdot q v\cdot (P+p_1) - q \cdot (P+p_1) )
-b_8 q\cdot p_1
-\frac{g_{\pi N \Delta}}{4 m} (v\cdot P v\cdot p_1 - P \cdot p_1 ) \Big]
\no\\&&
+\frac{P_\mu}{m} \Big[
\frac{b_3}{2} z ((v\cdot q)^2 -q^2) +(b_3+b_8) (v\cdot q)^2\no\\&&
\hspace{0.7cm}-\frac{g_{\pi N\Delta}}{3 m} (-\frac{1}{2} ( v\cdot P v\cdot q - P\cdot q) -(2 z_0-1) v\cdot P v\cdot q 
                              +(2 z_0-4) \Delta v\cdot q)\Big]
\Bigg\}~.
\eea
Similarly, for the case of an 
outgoing nucleon (momentum $p_2$), an incoming delta (momentum $P_\mu$, isospin $i$), 
and an outgoing pion (momentum $q$, isospin $a$), we have:\\
\underline{1st order:}
\be
\frac{g_{\pi N \Delta}}{F} q^\mu \delta^{ia}~.
\ee
\underline{2nd order:}
\be
{\bf +}\frac{b_3+b_8}{F} v\cdot q q^\mu \delta^{ia}
-\frac{g_{\pi N \Delta}}{m F} P_\mu v\cdot q \delta^{ia}~.
\ee
\underline{3rd order:}
\bea
\delta^{ia} \frac{1}{F} & \Bigg\{
&- q_\mu \Big( (e_1+e_2) (v\cdot q)^2 +2 M^2 (-2 e_4 +e_5) \Big)
\no\\&&
+\frac{q_\mu}{m} \Big[
-\frac{b_3}{2} (v\cdot q v\cdot (P+p_2) - q \cdot (P+p_2) )
+b_8 q\cdot p_2
-\frac{g_{\pi N \Delta}}{4 m} (v\cdot P v\cdot p_2 - P \cdot p_2 ) \Big]
\no\\&&
+\frac{P_\mu}{m} \Big[
-\frac{b_3}{2} z ((v\cdot q)^2 -q^2) -(b_3+b_8) (v\cdot q)^2\no\\&&
\hspace{0.7cm}-\frac{g_{\pi N\Delta}}{3 m} (-\frac{1}{2} ( v\cdot P v\cdot q - P\cdot q) -(2 z_0-1) v\cdot P v\cdot q 
                              +(2 z_0-4) \Delta v\cdot q)\Big]
\Bigg\} ~.
\eea
Note that possible $\pi \pi N \Delta$ vertices only start at second
order and can therefore not appear in the leading loop graphs of order
${\cal O}(\ve^3)$. This explains the vanishing of some diagrams not
drawn in Fig.\ref{fig:loop}.

\section{Expressions for the $1/m$ corrections}
\setcounter{equation}{0}
\label{app:1/m}

In this appendix, we give the lengthy  expressions for the
various $1/m$ corrections stemming from the $\Delta$--insertions,
which are of relevance to our problem.

\medskip\noindent
\underline{$1/m$ corrections to ${\cal L}_{\pi N}^{(2)}$:}

\bea
\bar{N} \gamma_0 {\cal B}_{N \Delta }^{\dagger (1)} \gamma_0 {\cal C}_\Delta^{-1(0) }
{\cal B}_{N \Delta }^{(1)} N
= 
-\frac{g_{\pi N \Delta}^2}{2 m} \bar{N} &\Big\{ &
\frac{4}{3} (1+8 z_0 + 12 z_0^2) S\cdot w^i \xi^{ij} S\cdot w^j\no\\&
+&\frac{1}{3} (5-8 z_0 -4 z_0^2) v\cdot w^i \xi^{ij} v\cdot w^j
\Big\} N ~.
\eea
\medskip\noindent
\underline{$1/m$  corrections to ${\cal L}_{\pi N}^{(3)}$:}

\bea
\bar{N} & \Big\{& 
[\gamma_0 ({\cal C}_{N \Delta}^{(1)} {\cal C}_\Delta^{-1 (0)} {\cal B}_{N \Delta}^{(1)} ) ^{-1}
\gamma_0 {\cal C}_N^{(0) -1} {\cal B}_N +{\rm h.c.}] 
+\gamma_0 {\cal B}_{N \Delta}^{\dagger (1)} \gamma_0 {\cal C}_{\Delta}^{(1) -1} 
{\cal B}_{N \Delta}^{(1)} \no\\
&+&[\gamma_0 {\cal B}_{N \Delta}^{\dagger (2)} \gamma_0 {\cal C}_{\Delta}^{(0) -1} 
{\cal B}_{N \Delta}^{(1)} +\rm h.c.] \Big\} N 
\no\\
=\bar{N} & \Big\{&\frac{\gpnd^2}{(2 m)^2}
\Big[ \frac{2}{3} (1+4 z_0 + 12 z_0^2) S\cdot w^i \xi^{ij} v\cdot w^j 2 i S\cdot D 
             + {\rm h.c.} \no\\&&
-\frac{2}{3} (3 + 4 z_0 + 4 z_0^2) v\cdot w^i \xi^{ij} S\cdot w^j 2 i S\cdot D 
             + {\rm h.c.}
-\frac{8}{3} (1+4 z_0 + 4 z_0^2) S\cdot w^i \xi^{ij} i v\cdot D S\cdot w^j\no\\&&
+\frac{16}{3} (1 + 6 z_0 + 8 z_0^2) \Delta S\cdot w^i \xi^{ij} S\cdot w^j
+\frac{8}{9} (5 +8 z_0 +12 z_0^2) S\cdot w^i \xi^{ij} i S\cdot D v\cdot w^j +{\rm h.c.}
\no\\&&
+\frac{2}{3} (-1 +4 z_0 -4 z_0^2) v\cdot w^i i v\cdot D \xi^{ij} v\cdot w^j
+4 (1-2 z_0) \Delta v\cdot w^i \xi^{ij} v\cdot w^j \Big]\no\\
&+&\frac{\gpnd}{2m} i 
\Big[ -\frac{4}{3} (1 +4 z +4 z_0 + 12 z z_0) (b_3 + b_8) 
S\cdot w^i \xi^{ij} w_{\alpha \beta}^j S_\alpha v_\beta - {\rm h.c.}\no\\&&
+\frac{4}{3} (1+4 z + 12 z z_0) b_3
S\cdot w^i \xi^{ij} w_{\alpha \beta}^j v_\alpha S_\beta - {\rm h.c.}\no\\&&
-\frac{2}{3}(6+13 z + 4 z z_0)b_3
v\cdot w^i \xi^{ij} w_{\alpha\beta}^j S_\alpha S_\beta - {\rm h.c.}\no\\&&
+\frac{1}{3} (-5 +4 z + 4 z_0 + 4 z z_0) (b_3 + b_8)
v\cdot w^i \xi^{ij} w_{\alpha \beta}^j v_\alpha v_\beta - {\rm h.c.}
\Big] \Big\} N ~.
\eea
\medskip\noindent
\underline{$1/m$  corrections to ${\cal L}_{\pi N \Delta}^{(2)}$:}

\bea
\bar{T}_\mu^i \gamma_0 {\cal B}_\Delta^{\dagger (1)} {\cal C}_\Delta^{-1 (0) }
{\cal B}_{N \Delta} ^{(1)} N + {\rm h.c.}
= -\frac{\gpnd}{2m} 2 \bar{T}_\mu^i  i D_\mu^{ik} \xi^{kl} v\cdot w^l N+ {\rm h.c.} ~.
\eea
\medskip\noindent
\underline{$1/m$  corrections to ${\cal L}_{\pi N \Delta}^{(3)}$:}

\bea
&&\bar{T}_\mu^i \Big\{ 
\gamma_0 \tilde{{\cal D}}_{N \Delta}^{\dagger (2)} \gamma_0 
\tilde{{\cal C}}_N^{-1 (0)} \tilde{{\cal B}}_N^{(1)}
+\gamma_0 {\cal B}_{\Delta}^{\dagger (1)} \gamma_0 
{\cal C}_\Delta^{-1 (1)} {\cal B}_{N \Delta}^{(1)}
+\gamma_0 {\cal B}_{\Delta}^{\dagger (1)} \gamma_0 
{\cal C}_\Delta^{-1 (0)} {\cal B}_{N \Delta}^{(2)} \Big\} N+ {\rm h.c.}\no\\
&=& \bar{T}_\mu^i \Big\{
\frac{-2 i b_3}{2m} S_\beta w_{\beta \mu}^i 2 i S\cdot D 
-\frac{4 i b_3}{2m}  i S\cdot D^{ik} \xi^{kl} w_{\mu\beta}^{l} S_\beta
-\frac{2 i (b_3+b_8)}{2m}i D_\mu^{ik} \xi^{kl} w_{\alpha\beta}^l v_\alpha v_\beta
\no\\&&
-\frac{\gpnd}{(2m)^2} 
\Big[2 i S\cdot D^{ik} \xi^{kj} w_\mu^j 2 i S\cdot D
-\frac{8+32 z_0}{3} i S\cdot D^{ik} \xi^{kl} i D_\mu^{lm} \xi^{mn} S\cdot w^n\no\\&&
+\frac{32}{3} z_0 i D_\mu^{ik} \xi^{kl} i S\cdot D^{lm} \xi^{mn} S\cdot w^n
-\frac{8 z_0-4}{3} i D_\mu^{ik} \xi^{kl} i v\cdot D^{lm} \xi^{mn} v\cdot w^n\no\\&&
+\frac{8 z_0-16}{3}\Delta i D_\mu^{ik} \xi^{kl}v\cdot w^l
\Big] \Big\} N+ {\rm h.c.} ~.
\eea

\medskip\noindent
\underline{$1/m$ corrections to ${\cal L}_{\pi \Delta}^{(2)}$:}

\bea
\bar{T}_\mu^i \gamma_0 {\cal B}_\Delta^{\dagger (1)} \gamma_0 {\cal
C}_\Delta^{-1 (0)}
{\cal B}_\Delta^{(1)}T_\nu^j
& = & \frac{1}{2m} \bar{T}_\mu^i \delta^{ij} 
2 i S\cdot D g_{\mu\nu} 2 i S\cdot T_\nu^j ~.
\eea

\medskip\noindent
\underline{$1/m$ corrections to ${\cal L}_{\pi \Delta}^{(3)}$:}

\bea
\bar{T}_\mu^i \gamma_0 {\cal B}_\Delta^{\dagger (1)} \gamma_0 {\cal
C}_\Delta^{-1 (1)}
{\cal B}_\Delta^{(1)}T_\nu^j
& = -\frac{1}{(2m)^2} \bar{T}_\mu^i \delta^{ij} & \Big\{ 
2 i S\cdot D g_{\mu\nu} (i v\cdot D + \Delta ) 2 i S\cdot D \no\\&&
+ 4 i D_\mu (i v\cdot D - \Delta) i D_\nu \Big\} T_\nu^j ~.
\eea

\section{Tree and loop amplitudes}
\setcounter{equation}{0}
\label{app:amp}

In this appendix, we give the lengthy analytical expressions for the
tree and counterterm as well as one loop amplitudes involving intermediate delta states, as
depicted in Fig.\ref{fig:tree} (tree and counterterm graphs) and Fig.\ref{fig:loop} (loop
graphs).  We use the loop functions defined in
Ref.\cite{moj}. All other notation has been defined previously.

\medskip

\noindent \underline{Tree diagram contributions:}

\begin{eqnarray}
F_\pi^2 g^+(\w,t) & = &
- \gpnd^2 \frac{2}{9}  (2 \w^2 -2 M_\pi^2 +t) \left( \frac{1}{\w-\de} -\frac{1}{\w+\de} \right) \no\\
&+& \gpnd^2 \frac{1}{9m} \Big\{ (2 \w^2 - 2 M_\pi^2 +t) (\w^2-M_\pi^2) \left( \left(\frac{1}{\w-\de} \right)^2+\left(\frac{1}{\w+\de} \right)^2\right) \no\\&&
-(2 \w^2 -2 M_\pi^2 +t) (4 \w^2 -4 M_\pi^2 +t) \left(\frac{1}{\w+\de} \right)^2 \no\\&&
+4 \w (4 \w^2 -4 M_\pi^2 +t) \frac{1}{\w+\de} \no\\&&
-4 \w^2 (3+4 z_0^2) +(2 M_\pi^2-t) (1+8 z_0 +12 z_0^2) \Big\}\no\\
&-&\gpnd (b_3+b_8) \frac{4}{9} \w (2 \w^2 - 2 M_\pi^2 +t) \left( \frac{1}{\w-\de} +\frac{1}{\w+\de} \right) \no\\
&+&\gpnd^2 \frac{1}{9m^2} \Big\{
-\frac{1}{2} (2 \w^2-2 M_\pi^2 +t) (\w^2-M_\pi^2)^2 \left( \left(\frac{1}{\w-\de} \right)^3-\left(\frac{1}{\w+\de} \right)^3\right) \no\\&&
+\frac{1}{2} (2 \w^2-2 M_\pi^2 +t)^2 (4 \w^2 -4 M_\pi^2 +t) \left(\frac{1}{\w+\de} \right)^3 \no\\&&
-\w (4 \w^2 -4 M_\pi^2 +t) (8 \w^2-8 M_\pi^2 +3 t)\left(\frac{1}{\w+\de} \right)^2 \no\\&&
+\frac{1}{6} (4 \w^2-4 M_\pi^2 +t) (72 \w^2 -24 M_\pi^2 +7 t )\frac{1}{\w+\de} \no\\&&
+16 \de \w^2 (1+2 z_0^2) -2 \de (2 M_\pi^2-t) (1+6 z_0 +8 z_0^2) - 2 \w (4 \w^2-4 M_\pi^2 +t) (3+4 z_0^2) \Big\}\no\\
&+&\gpnd (b_3+b_8) \frac{1}{9 m} \Big\{
2 \w (\w^2-M_\pi^2) (2\w^2-2 M_\pi^2+t) \left( \left(\frac{1}{\w-\de} \right)^2-\left(\frac{1}{\w+\de} \right)^2\right) \no\\&&
+2 \w (2 \w^2-2 M_\pi^2 +t) (4\w^2 -4 M_\pi^2 +t) \left(\frac{1}{\w+\de} \right)^2 \no\\&&
-2[ (\w^2-M_\pi^2) (22 \w^2 -6 M_\pi^2 +5 t) +t^2 ] \frac{1}{\w+\de}\no\\&&
-2 (\w^2-M_\pi^2) (2 \w^2-2 M_\pi^2 +t) \frac{1}{\w-\de} \Big\}\no\\
&-&\gpnd b_8 \frac{2}{9 m} (2 \w^2-2 M_\pi^2 +t) (\w^2-M_\pi^2) \left( \frac{1}{\w-\de} -\frac{1}{\w+\de} \right)\no\\
&+&[2 \gpnd (e_1+e_2) - (b_3+b_8)^2] \frac{2}{9} \w^2 (2 \w^2-2 M_\pi^2 +t) \left( \frac{1}{\w-\de} -\frac{1}{\w+\de} \right)\no\\
&-&\gpnd (2 e_4 -e_5) \frac{8}{9} M_\pi^2  (2 \w^2-2 M_\pi^2 +t) \left( \frac{1}{\w-\de} -\frac{1}{\w+\de} \right)~.\\
F_\pi^2 h^+(\w,t) & = &
\gpnd^2 \frac{2}{9} \left( \frac{1}{\w-\de} +\frac{1}{\w+\de} \right)\no\\
&-& \gpnd^2 \frac{1}{9m} \Big\{  (\w^2-M_\pi^2) \left( \left(\frac{1}{\w-\de} \right)^2-\left(\frac{1}{\w+\de} \right)^2\right) \no\\&&
+ (4 \w^2-4 M_\pi^2 +t) \left(\frac{1}{\w+\de} \right)^2\no\\&&
-4 \w \frac{1}{\w+\de} \Big\}\no\\
&+&\gpnd (b_3 + b_8) \frac{4}{9}\w \left( \frac{1}{\w-\de} -\frac{1}{\w+\de} \right)\no\\
&+&\gpnd^2 \frac{1}{9 m^2} \Big\{ \frac{1}{2}(\w^2 - M_\pi^2)^2 \left( \left(\frac{1}{\w-\de} \right)^3+\left(\frac{1}{\w+\de} \right)^3\right) \no\\&&
+\frac{1}{2} (2\w^2-2 M_\pi^2 +t) (4\w^2 -4 M_\pi^2 +t) \left(\frac{1}{\w+\de} \right)^3\no\\&&
- \w (10 \w^2-10 M_\pi^2 +3 t) \left(\frac{1}{\w+\de} \right)^2\no\\&&
+ (16 \w^2 -12 M_\pi^2 +3 t) \frac{1}{\w+\de} \no\\&&
-2 \w (1+4 z_0^2) \Big\}\no\\
&+&\gpnd (b_3 + b_8) \frac{1}{9m} \Big\{ -2 \w (\w^2-M_\pi^2) \left( \left(\frac{1}{\w-\de} \right)^2+\left(\frac{1}{\w+\de} \right)^2\right) \no\\&&
+2 \w (4 \w^2-4 M_\pi^2 +t) \left(\frac{1}{\w+\de} \right)^2\no\\&&
-2 (7\w^2 -3 M_\pi^2 +t) \frac{1}{\w+\de}\no\\&&
+2 (\w^2-M_\pi^2) \frac{1}{\w-\de} \Big\} \no\\&&
-16 \w z_0 \no\\
&+&\gpnd b_8 \frac{2}{9m} \Big\{ (\w^2-M_\pi^2) \left( \frac{1}{\w-\de} +\frac{1}{\w+\de} \right)
-2 \w (1+4 z +12 z z_0) \Big\}\no\\
&-&[ 2\gpnd (e_1+e_2) -(b_3+b_8)^2] \frac{2}{9} \w^2 \left( \frac{1}{\w-\de} +\frac{1}{\w+\de} \right)\no\\
&+&\gpnd (2 e_4 - e_5) \frac{8}{9} M_\pi^2 \left( \frac{1}{\w-\de} +\frac{1}{\w+\de} \right)~.\\
F_\pi^2 g^-(\w,t) & = &
\gpnd^2 \frac{1}{9} (2 \w^2-2 M_\pi^2 +t) \left( \frac{1}{\w-\de} +\frac{1}{\w+\de} \right)\no\\
&+& \gpnd^2 \frac{1}{9m} \Big\{ 
-\frac{1}{2} (2 \w^2 -2 M_\pi^2 +t) (\w^2-M_\pi^2)  \left( \left(\frac{1}{\w-\de}\right)^2 -\left(\frac{1}{\w+\de} \right)^2 \right)\no\\&&
-\frac{1}{2} (2 \w^2-2 M_\pi^2 +t) (4 \w^2-4 M_\pi^2 +t) \left(\frac{1}{\w+\de} \right)^2\no\\&&
+2 \w (4 \w^2-4 M_\pi^2 +t) \frac{1}{\w+\de} \Big\}\no\\
&+&\gpnd (b_3+b_8) \frac{2}{9} w (2 \w^2-2 M_\pi^2 +t)  \left( \frac{1}{\w-\de} -\frac{1}{\w+\de} \right)\no\\
&+&\gpnd^2 \frac{1}{9 m^2} \Big\{
\frac{1}{4} (2 \w^2-2 M_\pi^2 +t) (\w^2-M_\pi^2)^2  \left( \left(\frac{1}{\w-\de}\right)^3 +\left(\frac{1}{\w+\de} \right)^3 \right)\no\\&&
+\frac{1}{4} (2\w^2-2M_\pi^2+t)^2 (4\w^2-4 M_\pi^2+t) \left(\frac{1}{\w+\de} \right)^3 \no\\&&
-\frac{1}{2} w (4\w^2-4 M_\pi^2 +t) (8\w^2-8M_\pi^2+3t) \left(\frac{1}{\w+\de} \right)^2 \no\\&&
+\frac{1}{12} (4\w^2-4M_\pi^2+t) (72\w^2-24 M_\pi^2 +7 t) \frac{1}{\w+\de}\no\\&&
+w [2 (2 M_\pi^2-t)(1-2 z_0 -2 z_0^2)-12 \w^2 + M_\pi^2 (9 +4 z_0 +12 z_0^2)] \Big\}\no\\
&+&\gpnd (b_3+b_8) \frac{1}{9m} \Big\{
-w (2\w^2-2 M_\pi^2+t) (\w^2-M_\pi^2) \left( \left(\frac{1}{\w-\de}\right)^2 +\left(\frac{1}{\w+\de} \right)^2 \right)\no\\&&
+ w (2\w^2-2 M_\pi^2+t) (4\w^2-4M_\pi^2+t) \left(\frac{1}{\w+\de} \right)^2 \no\\&&
+[(\w^2-M_\pi^2) (-22\w^2+6 M_\pi^2 -5t) -t^2] \frac{1}{\w+\de}\no\\&&
+(2\w^2-2 M_\pi^2+t) (\w^2-M_\pi^2) \frac{1}{\w-\de}\no\\&&
-4 \w (2 M_\pi^2-t) z_0 +\w^3 (16 +5 z -4 z z_0) - \w M_\pi^2 (6 +13 z +4 z z_0) \Big\} \no\\
&+&\gpnd b_8 \frac{1}{9m} \Big\{
(2\w^2-2 M_\pi^2+t) (\w^2-M_\pi^2) \left( \frac{1}{\w-\de}+\frac{1}{\w+\de}  \right)\no\\&&
-\w (2 M_\pi^2-t) (1+4 z +12 z z_0) -\w^3 (4+5 z -20 z z_0) +\w M_\pi^2 (6+13 z +4 z z_0) \Big\} \no\\
&-&[ 2\gpnd (e_1+e_2) -(b_3+b_8)^2]  \frac{1}{9} \w^2 (2\w^2-2 M_\pi^2+t)  \left( \frac{1}{\w-\de}+\frac{1}{\w+\de}  \right)\no\\
&+&\gpnd (2 e_4-e_5) \frac{4}{9} M_\pi^2 (2\w^2-2 M_\pi^2+t)  \left( \frac{1}{\w-\de}+\frac{1}{\w+\de}  \right)~.\\
F_\pi^2 h^-(\w,t) & = &
-\gpnd^2 \frac{1}{9}  \left( \frac{1}{\w-\de}-\frac{1}{\w+\de}  \right)\no\\
&+&\gpnd^2 \frac{1}{9m} \Big\{\frac{1}{2}(\w^2-M_\pi^2)  \left( \left(\frac{1}{\w-\de}\right)^2 +\left(\frac{1}{\w+\de} \right)^2 \right)\no\\&&
-\frac{1}{2}(4\w^2-4M_\pi^2+t) \left(\frac{1}{\w+\de} \right)^2 \no\\&&
+2 \w \frac{1}{\w+\de} \no\\&&
+ (1+8 z_0 +12 z_0^2)\Big\}\no\\
&-&\gpnd (b_3+b_8) \frac{2}{9} \w  \left( \frac{1}{\w-\de}+\frac{1}{\w+\de}  \right)\no\\
&+&\gpnd^2 \frac{1}{9m^2} \Big\{
-\frac{1}{4} (\w^2-M_\pi^2)^2 \left( \left(\frac{1}{\w-\de}\right)^3 -\left(\frac{1}{\w+\de} \right)^3 \right)\no\\&&
+\frac{1}{4} (2\w^2-2 M_\pi^2+t) (4\w^2-4M_\pi^2+t) \left(\frac{1}{\w+\de} \right)^3 \no\\&&
-\frac{1}{2}\w (10 \w^2-10 M_\pi^2 +3 t)  \left(\frac{1}{\w+\de} \right)^2 \no\\&&
+\frac{1}{2}(16\w^2 -12 M_\pi^2 +3t) \frac{1}{\w+\de} \no\\&&
+\w (1+8 z_0 +12 z_0^2)-2 \de (1+6z_0 +8 z_0^2) \Big\}\no\\
&+&\gpnd (b_3+b_8) \frac{1}{9m} \Big\{
\w (\w^2-M_\pi^2)  \left( \left(\frac{1}{\w-\de}\right)^2 -\left(\frac{1}{\w+\de} \right)^2 \right)\no\\&&
+ \w (4\w^2-4 M_\pi^2+t) \left(\frac{1}{\w+\de} \right)^2 \no\\&&
- (7\w^2-3 M_\pi^2+t) \frac{1}{\w+\de}\no\\&&
- (\w^2-M_\pi^2) \frac{1}{\w-\de}\no\\
&-&\gpnd b_8 \frac{1}{9m} (\w^2-M_\pi^2)  \left( \frac{1}{\w-\de} -\frac{1}{\w+\de} \right)\no\\
&+&[ 2\gpnd (e_1+e_2) -(b_3+b_8)^2]  \frac{1}{9} \w^2   \left( \frac{1}{\w-\de}-\frac{1}{\w+\de}  \right)\no\\
&-&\gpnd (2e_4-e_5)   \frac{4}{9} M_\pi^2   \left( \frac{1}{\w-\de}-\frac{1}{\w+\de}  \right)~.
\end{eqnarray}

\medskip

\noindent\underline{Loop diagram contributions:}

\begin{eqnarray}
F_\pi^4 g^+(\w,t) & = & 
(J_0(\w)+J_0(\w))\Big\{ 
-\frac{8\gpnd^2 g_A^2 }{27 (\w^2-\de^2)} (2 M_\pi^4 -4 \w^2 M_\pi^2 -t M_\pi^2 + 2 \w^4 + \w^2 t )\no\\&&\hspace{1cm}
+\frac{8 \gpnd^4}{243 (\w^2-\de^2)^2}  ( 8 M_\pi^4 \w^2 + 10 M_\pi^4 \de^2 -16 M_\pi^2 \w^4 -4 M_\pi^2 \w^2 t \no\\&&\hspace{2cm}
   - 20 M_\pi^2 \w^2 \de^2 -5 M_\pi^2 t \de^2 
   +8 \w^6+4 \w^4 t +10 \w^4 \de^2 +5 \w^2 t \de^2)\Big\}\no\\
&+&J_0(-\de)\Big\{
\frac{\gpnd^2 }{2187\w^2 (\w^2-\de^2)^2} (-972 M_\pi^2 \w^6 -972 M_\pi^2 \w^2 \de^4 + 1944 M_\pi^2 \w^4 \de^2 \no\\&&\hspace{2cm}
   -3888 \w^4 t \de^2 +1944 \w^2 t \de^4 + 1944 \w^6 t)\no\\&&\hspace{1cm}
+\frac{\gpnd^4 }{2187\w^2 (\w^2-\de^2)^2} (-792 M_\pi^4 \w^2 \de^2-504 M_\pi^4 \w^4 + 7776 M_\pi^2 \w^4 \de^2 \no\\&&\hspace{2cm}
   + 252 M_\pi^2 \w^4 t +396 M_\pi^2 \w^2 t \de^2 -5688 M_\pi^2 \w^2 \de^4 +504 M_\pi^2 \w^6 \no\\&&\hspace{2cm}
   -3492 \w^4 t \de^2 +2844 \w^2 t \de^4 +5688 \w^4 \de^4-6984 \w^6 \de^2) \no\\&&\hspace{1cm}
+\frac{\gpnd^2 g_1^2 }{2187\w^2 (\w^2-\de^2)^2} ( -50 M_\pi^4 \w^4 +500 M_\pi^4 \w^2 \de^2 -450 M_\pi^4 \de^4 +50 M^2 \w^6\no\\&&\hspace{2cm}
   -450 M_\pi^2 \w^4 \de^2 +225 M_\pi^2 t \de^4 +25 M_\pi^2 \w^4 t -250 M_\pi^2 \w^2 t \de^2 -50 M_\pi^2 \w^2 \de^4 \no\\&&\hspace{2cm}
   +450 M_\pi^2 \de^6+ 500 \w^4 \de^4 -225 t \de^6 +250 \w^2 t \de^4 -50 \w^6 \de^2 -450 \w^2 \de^6 \no\\&&\hspace{2cm}
   -25 \w^4 t \de^2)\no\\&&\hspace{1cm}
-\frac{50 \gpnd^2 g_A g_1}{243 \w^2} (-2 M_\pi^4 + 2 \w^2 M_\pi^2 + M_\pi^2 t +2 M_\pi^2 \de^2 -2 \w^2 \de^2 -t \de^2 )\no\\&&\hspace{1cm}
+\frac{\gpnd^2 g_A^2}{27 \w^2 (\w^2-\de^2)}  (-2 M_\pi^4 \w^2 +18 M_\pi^4 \de^2 +2 M_\pi^2 \w^4 + M_\pi^2 \w^2 t  -16 M_\pi^2 \w^2 \de^2 
       \no\\&&\hspace{2cm}
   -18 M_\pi^2 \de^4 -9 M_\pi^2 t \de^2 -2 \w^4 \de^2 -\w^2 t \de^2 +18 \w^2 \de^4 +9 t \de^4) \Big\}\no\\
&-&J_0(\de)\frac{4 \gpnd^4}{27 (\w^2-\de^2)^2} (-2 M_\pi^4 \w^2 -2 M_\pi^4 \de^2 + M_\pi^2 t \de^2 +4 M_\pi^2 \w^2 \de^2 +2 M_\pi^2 \de^4 
      \no\\&&\hspace{2cm}
   +2 M_\pi^2 \w^4 + M_\pi^2 \w^2 t-t \de^4 -2 \w^4 \de^2 -2 \w^2 \de^4 -\w^2 t \de^2 )\no\\
&+&\frac{1}{\w^2} \left(\frac{J_0(\w-\de)}{(\w-\de)^2} + \frac{J_0(-\w-\de)}{(\w+\de)^2}\right)\Big\{
\frac{25 \gpnd^2 g_1^2}{4374} (20 M_\pi^4 \w^2 +18 M_\pi^4 \de^2 -40 M_\pi^2 \w^4 \no\\&&\hspace{2cm}
   -10 M_\pi^2 \w^2 t -96 M_\pi^2 \w^2 \de^2 -9 M_\pi^2 t \de^2 -18 M_\pi^2 \de^4  +20 \w^6 +78 \w^4 \de^2 \no\\&&\hspace{2cm}
   +10 \w^4 t +39 \w^2 t \de^2 +18 \w^2 \de^4 +9 t \de^4)\no\\&&\hspace{1cm}
-\frac{25 \gpnd^2 g_A g_1}{243} ( 4 M_\pi^4 \w^2 +2 M_\pi^4 \de^2-8 M_\pi^2 \w^4 -2 M_\pi^2 \w^2 t -16 M_\pi^2 \w^2 \de^2 \no\\&&\hspace{2cm}
   -2 M_\pi^2 \de^4 -M_\pi^2 t \de^2 +4 \w^6 +14 \w^4 \de^2 +7 \w^2 t \de^2 +2 \w^2 \de^4 +t \de^4 +2 \w^4 t)\no\\&&\hspace{1cm}
+\frac{\gpnd^2 g_A^2}{54} ( 20 M_\pi^4 \w^2 + 18 M_\pi^4 \de^2 -40 M_\pi^2 \w^4 -10 M_\pi^2 \w^2 t -96 M_\pi^2 \w^2 \de^2 \no\\&&\hspace{2cm}
   -9 M_\pi^2 t \de^2 -18 M_\pi^2 \de^4 +20 \w^6 +78 \w^4 \de^2 +10 \w^4 t +39 \w^2 t \de^2  \no\\&&\hspace{2cm}
   +18 \w^2 \de^4 +9 t \de^4) \Big\}\no\\
&+&\frac{\de}{\w} \left(\frac{J_0(\w-\de)}{(\w-\de)^2} - \frac{J_0(-\w-\de)}{(\w+\de)^2}\right)\Big\{
\frac{25 \gpnd^2 g_1^2}{4374} (-20 M_\pi^4 +80 M_\pi^2 \w^2  \no\\&&\hspace{2cm}
   +56 M_\pi^2 \de^2 +10 M_\pi^2 t  -60 \w^4  -56 \w^2 \de^2 -30 \w^2 t  -28 t \de^2)\no\\&&\hspace{1cm}
-\frac{25 \gpnd^2 g_A g_1}{243} (-4 M_\pi^4  +16 M_\pi^2 \w^2  +8 M_\pi^2 \de^2 +2 M_\pi^2 t -12 \w^4 -6 \w^2 t  \no\\&&\hspace{2cm}
   -8 \w^2 \de^2 -4 t \de^2 )\no\\&&\hspace{1cm}
+\frac{\gpnd^2 g_A^2}{54} (-20 M_\pi^4  +80 M_\pi^2 \w^2  +56 M_\pi^2 \de^2 +10 M_\pi^2 t  -60 \w^4  \no\\&&\hspace{2cm}
   -56 \w^2 \de^2 -30 \w^2 t  -28 t \de^2 ) \Big\}\no\\
&-&J_0^\prime(-\de) \frac{20 \gpnd^4}{27 (\w^2-\de^2)} \de (-2 M_\pi^4 +2 M_\pi^2 \w^2 + M_\pi^2 t +2 M_\pi^2 \de^2 -2 \w^2\de^2 -t \de^2)\no\\
&-& K_0(t,-\de) \frac{2 \gpnd^2}{9} ( 2 M_\pi^4 -2 M_\pi^2 \de^2 -5 M_\pi^2 t +2 t^2 +4 t \de^2)\no\\
&+& I_0(t) \frac{4 \gpnd^2}{9} \de (M_\pi^2 -2 t)\no\\
&+& \frac{\gpnd^4}{104976 \pi^2 (\w^2-\de^2)^2} (1296 M_\pi^4 \w^2 \de+10368 M_\pi^4 \de^3 -14688 M_\pi^2 \de^5 \no\\&&\hspace{1cm}
   -5184 M_\pi^2 t \de^3 -648 M_\pi^2 \w^2 t \de -3456 M_\pi^2 \w^2 \de^3 -1296 M_\pi^2 \w^4 \de +14688 \w^2 \de^5 \no\\&&\hspace{1cm}
   -6912 \w^4 \de^3 -3456 \w^2 t \de^3 +7344 t \de^5)\no\\&&
+\frac{\gpnd^2}{104976 \pi^2 (\w^2-\de^2)^2} \de ( 1944 M_\pi^4 \ (\w^2- \de^2) -2916 M_\pi^2 \de^4 +7776 M_\pi^2 \w^2 \de^2 \no\\&&\hspace{1cm}
   +972 M_\pi^2 t \de^2 -4860 M_\pi^2 \w^4 -972 M_\pi^2 \w^2 t +5832 t (\w^2-\de^2)^2)\no\\&&
+\frac{\gpnd^2 g_1^2}{104976 \pi^2 (\w^2-\de^2)^2} (-2700 \pi M_\pi^5 (\w^2+\de^2) -3030 M_\pi^4 \de (\w^2-\de^2) \no\\&&\hspace{1cm}
   +1350 \pi M_\pi^3 (\w^2+\de^2) (2 \w^2 +t) +1515 M_\pi^2 \w^2 t \de -9840 M_\pi^2 \de^5 \no\\&&\hspace{1cm}
   +18410 M_\pi^2 \w^2 \de^3 -1515 M_\pi^2 t \de^3 -8570M^2 \w^4 \de -21440 \w^4 \de^3 +9840 \w^2 \de^5 \no\\&&\hspace{1cm}
   +4920 t \de^5 +11600 \w^6 \de -10720 \w^2 t \de^3 +5800 \w^4 t \de)\no\\&&
-\frac{5 \gpnd^2 g_A g_1}{1944 \pi^2 (\w^2-\de^2)^2} ( 20 \pi M_\pi^5 (\w^2 + \de^2) -42 M_\pi^4 \de (\w^2-\de^2) \no\\&&\hspace{1cm}
   -10 \pi M_\pi^3 (\w^2 + \de^2) ( 2 \w^2 + t) -21 M_\pi^2 t \de^3 +21 M_\pi^2 \w^2 t \de -112 M_\pi^2 \de^5 \no\\&&\hspace{1cm}
   -70 M_\pi^2 \w^4 \de +182 M_\pi^2 \w^2 \de^3+112 \w^6 \de +56 t \de^5 +112 \w^2 \de^5 -112 \w^2 t \de^3 \no\\&&\hspace{1cm}
   -224 \w^4 \de^3 +56 \w^4 t \de)\no\\&&
+\frac{\gpnd^2 g_A^2}{648 \pi^2 (\w^2-\de^2)^2} ( -\pi M^5(42 \w^2 - 150 \de^2)-24 M^4 \de (\w^2- \de^2)\no\\&&\hspace{1cm}
   + \pi M^3 (21 \w^2-75 \de^2)(2 \w^2+ t) + 200 M^2\w^2 \de^3-80 M^2 \w^4 \de -120 M^2\de^5\no\\&&\hspace{1cm}
   +12 M^2\w^2 t \de -12 M^2 t \de^3 +104 \w^6 \de+52 t \w^4 \de -112 t \w^2 \de^3+60 t \de^5\no\\&&\hspace{1cm}
   -224 \w^4 \de^3+120 \w^2 \de^5)~.\\
F_\pi^4 h^+(\w,t) & = & 
(J_0(\w)+J_0(-\w))\frac{\w^2-M_\pi^2}{\w^2-\de^2}\Big\{ -\frac{32 \gpnd^4}{81  (\w^2-\de^2)}  \w \de  
   +\frac{4 \gpnd^2 g_A^2}{9 w} \de \Big\}\no\\
&+&(J_0(\w)-J_0(-\w))\frac{\w^2-M_\pi^2}{ \w^2-\de^2} \Big\{\frac{8 \gpnd^4}{243 (\w^2-\de^2)} (2 w^2 +\de^2) -\frac{4\gpnd^2 g_A^2}{27}\Big\}\no\\
&+& J_0(\de) \frac{8 \gpnd^4}{27 (\w^2-\de^2)^2} \w \de (M_\pi^2-\de^2)\no\\
&+& J_0(-\de) \Big\{
\frac{4 \gpnd^4 \de}{2187 \w (\w^2-\de^2)^2} ( -270 M_\pi^2 \w^2 +918 \w^4 -648 \w^2 \de^2)\no\\&&\hspace{1cm}
+\frac{4 \gpnd^2 g_1^2 \de}{2187 \w (\w^2-\de^2)^2} ( 50 M_\pi^2 (\w^2-\de^2) +125 \de^4 -200 \w^2 \de^2 +75 \w^4)\no\\&&\hspace{1cm}
-\frac{200 \gpnd^2 g_A g_1 }{243 \w} \de 
+\frac{4 \gpnd^2 g_A^2}{27 \w \de (\w^2-\de^2)} ( 8 M_\pi^2 \w^2 -4 M_\pi^2 \de^2 + w^2 \de^2 -5 \de^4) \Big\}\no\\
&+&\frac{1}{\w} \left(\frac{J_0(\w-\de)}{(\w-\de)^2} + \frac{J_0(-\w-\de)}{(\w+\de)^2}\right)\Big\{
\frac{25 \gpnd^2 g_1^2}{4374} (20 M_\pi^2 \de -30 \w^2 \de -32 \de^3)\no\\&&\hspace{1cm}
-\frac{25 \gpnd^2 g_A g_1}{243} ( 4 M_\pi^2 \de -6 \w^2 \de -8 \de^3)\no\\&&\hspace{1cm}
+\frac{\gpnd^2 g_A^2}{27} (10 M_\pi^2 \de -15 \w^2 \de -28 \de^3) \Big\}\no\\
&+&\frac{1}{\w^2} \left(\frac{J_0(\w-\de)}{(\w-\de)^2} - \frac{J_0(-\w-\de)}{(\w+\de)^2}\right)\Big\{
\frac{25 \gpnd^2 g_1^2}{4374} (-5 M_\pi^2 \w^2 -6 M_\pi^2 \de^2 +5 \w^4 \no\\&&\hspace{2cm}
   +51 \w^2 \de^2 +6 \de^4)\no\\&&\hspace{1cm}
-\frac{25 \gpnd^2 g_A g_1}{243} ( -M_\pi^2 \w^2 -2 M_\pi^2 \de^2 + \w^4 +11 \w^2 \de^2 +2 \de^4)\no\\&&\hspace{1cm}
+\frac{\gpnd^2 g_A^2}{54} (-5 M_\pi^2 \w^2 -18 M_\pi^2 \de^2 +5 \w^4 +63 \w^2 \de^2 +18 \de^4) \Big\}\no\\
&+& J_0^\prime(-\de) (M_\pi^2-\de^2) \Big\{
\frac{50 \gpnd^2 g_1^2}{729 \w}
+\frac{68 \w \gpnd^4}{81  (\w^2-\de^2)}
-\frac{100 \gpnd^2 g_A g_1}{243 \w}\no\\&&\hspace{1cm}
+\frac{2 \gpnd^2 g_A^2}{3 \w} \Big\}\no\\
&-&\frac{\gpnd^4 \w}{26244 \pi^2 (\w^2-\de^2)^2}\de^2(-2916 M_\pi^2 +1944\de^4)\no\\&&
-\frac{\gpnd^2 g_1^2 \w}{52488 \pi^2 (\w^2-\de^2)^2} ( 1350 \pi M_\pi^3 \de +645 M_\pi^2 (w^2 -\de^2) +1700 \w^4 +1860 \de^4 \no\\&&\hspace{1cm}
   -3560 \w^2 \de^2)\no\\&&
-\frac{\gpnd^2 g_A g_1 \w}{54 \pi^2 (\w^2-\de^2)} M_\pi^2\no\\&&
+\frac{5 \gpnd^2 g_A g_1\w}{972 \pi^2 (\w^2-\de^2)^2}(-10 \pi M_\pi^3 \de +3 M_\pi^2 (\w^2-\de^2) +12 (\w^2-\de^2)^2)\no\\&&
+\frac{\gpnd^2 g_A^2}{324 \pi^2 \w \de (\w^2-\de^2)^2} (\pi M_\pi^3 (48 \w^4 -93 \w^2 \de^2 +36 \de^4) -42 M_\pi^2 \w^2 \de (\w^2 - \de^2) \no\\&&\hspace{1cm}
   +4 \w^4 \de^3 -10 \w^2 \de^5 +6 \de \w^6)~.\\
F_\pi^4 g^-(\w,t) & = & 
(J_0(\w)+J_0(-\w)) \frac{1}{\w^2-\de^2} \Big\{
\frac{16 \gpnd^4}{81 (\w^2-\de^2)} \w \de (-2 M_\pi^4 +4 M_\pi^2 \w^2 + M_\pi^2 t - \w^2 t -2\w^4)\no\\&&\hspace{1cm}
-\frac{2 \gpnd^2 g_A^2 }{27} (2 M_\pi^4 -4 M_\pi^2 \w^2 -M_\pi^2 t +2 \w^4 +\w^2 t) \Big\} \no\\
&+&(J_0(\w)-J_0(-\w)) \frac{1}{\w^2-\de^2} \Big\{
\frac{4 \gpnd^4}{243 (\w^2-\de^2)}  (2 M_\pi^4 \de^2 +4 M_\pi^4 \w^2 -8 M_\pi^2 \w^4 \no\\&&\hspace{2cm}
   -4 M_\pi^2 \w^2 \de^2 -2 M_\pi^2 \w^2 t -M_\pi^2 t \de^2 +2 \w^4 \de^2 +4 \w^6 +2 \w^4 t +\w^2 t \de^2)\no\\&&\hspace{1cm}
-\frac{2 \gpnd^2 g_A^2 }{9 \w}\de ( -2 M_\pi^4 +4 M_\pi^2 \w^2 + M_\pi^2 t -2 \w^4 - \w^2 t) \Big\} \no\\
&+&J_0(\de) \frac{4 \gpnd^4}{27 (\w^2-\de^2)^2} \w \de (-2 M_\pi^4 +2 M_\pi^2 \w^2 + M_\pi^2 t +2 M_\pi^2 \de^2 -2 \w^2 \de^2 -t \de^2)\no\\
&+&J_0(-\de) \Big\{
\frac{2 \gpnd^2 g_1^2}{2187 \w (\w^2-\de^2)^2} \de ( -100 M_\pi^4 (\w^2-\de^2) +50 M_\pi^2 \w^2 t -250 M_\pi^2 \de^4 \no\\&&\hspace{2cm}
   -50 M_\pi^2 \w^4 -50 M_\pi^2 t \de^2 +300 M_\pi^2 \w^2 \de^2 +125 t \de^4 -400 \w^4 \de^2  \no\\&&\hspace{2cm}
   +75 \w^4 t +250 \w^2 \de^4 -200 \w^2 t \de^2 +150 \w^6)\no\\&&\hspace{1cm}
+\frac{2 \gpnd^4}{2187 \w (\w^2-\de^2)^2}\de ( 540 M_\pi^4 \w^2 +1296 M_\pi^2 \w^2 \de^2 -270 M_\pi^2 \w^2 t \no\\&&\hspace{2cm}
   -2376 M_\pi^2 \w^4 +918 \w^4 t -648 \w^2 t \de^2 -1296 \w^4 \de^2 +1836 \w^6)\no\\&&\hspace{1cm}
-\frac{100 \gpnd^2 g_A g_1}{243 w} \de (-2 M_\pi^2 +2 \w^2 +t)\no\\&&\hspace{1cm}
+\frac{2 \gpnd^2 g_A^2}{27 \w \de (\w^2-\de^2)} ( -16 M_\pi^4 \w^2 +8 M_\pi^4 \de^2 +16 M_\pi^2 \w^4 +8 M_\pi^2 \w^2 t +10 M_\pi^2 \de^4 
      \no\\&&\hspace{2cm}
   -10 M_\pi^2 \w^2 \de^2 -4 M_\pi^2 t \de^2 +2 \w^4 \de^2 +\w^2 t \de^2 -10 \w^2 \de^4 -5 t \de^4) \Big\} \no\\
&+&\frac{\de}{\w} \left(\frac{J_0(\w-\de)}{(\w-\de)^2} + \frac{J_0(-\w-\de)}{(\w+\de)^2}\right)\Big\{
\frac{25 \gpnd^2 g_1^2}{8748}  (-40 M_\pi^4 +100 M_\pi^2 \w^2 +64 M_\pi^2 \de^2 \no\\&&\hspace{2cm}
   +20 M_\pi^2 t -60 \w^4 -30 \w^2 t -64 \w^2 \de^2 -32 t \de^2)\no\\&&\hspace{1cm}
-\frac{25  \gpnd^2 g_A g_1}{486}  (-8 M_\pi^4 +20 M_\pi^2 \w^2 +4 M_\pi^2 t +16 M_\pi^2 \de^2 -12 \w^4 -16 \w^2 \de^2 \no\\&&\hspace{2cm}
   -6 \w^2 t -8 t \de^2)\no\\&&\hspace{1cm}
+\frac{\gpnd^2 g_A^2}{108} (-40 M_\pi^4 +100 M_\pi^2 \w^2 +112 M_\pi^2 \de^2 +20 M_\pi^2 t -60 \w^4 -112 \w^2 \de^2 \no\\&&\hspace{2cm}
   -30 \w^2 t -56 t \de^2 ) \Big\}\no\\
&+&\frac{1}{\w^2} \left(\frac{J_0(\w-\de)}{(\w-\de)^2} - \frac{J_0(-\w-\de)}{(\w+\de)^2}\right)\Big\{
\frac{25 \gpnd^2 g_1^2}{8748} (10 M_\pi^4 \w^2 +12 M_\pi^4 \de^2 -20 M_\pi^2 \w^4 \no\\&&\hspace{2cm}
   -114 M_\pi^2 \w^2 \de^2 -5 M_\pi^2 \w^2 t -6 M_\pi^2 t \de^2 - 12 M_\pi^2 \de^4 +10 \w^6 +5 \w^4 t \no\\&&\hspace{2cm}
   +102 \w^4 \de^2 +51 \w^2 t \de^2 +12 \w^2 \de^4 +6 t \de^4)\no\\&&\hspace{1cm}
-\frac{25 \gpnd^2 g_A g_1}{486} ( 2 M_\pi^4 \w^2 +4 M_\pi^4 \de^2 -4 M_\pi^2 \w^4 -26 M_\pi^2 \w^2 \de^2 -M_\pi^2 w^2 t \no\\&&\hspace{2cm}
   -2 M_\pi^2 t \de^2 -4 M_\pi^2 \de^4 +2 \w^6 +\w^4 t +22 \w^4 \de^2 +4 \w^2 \de^4 +11 \w^2 t \de^2  \no\\&&\hspace{2cm}
   +2 t \de^4)\no\\&&\hspace{1cm}
+\frac{\gpnd^2 g_A^2}{108} ( 10 M_\pi^4 \w^2 +36 M^4 \de^2 -20 M_\pi^2 \w^4 -5 M_\pi^2 \w^2 t -162 M_\pi^2 \w^2 \de^2 \no\\&&\hspace{2cm}
   -18 M_\pi^2 t \de^2 -36 M_\pi^2 \de^4 +10 \w^6 +126 \w^4 \de^2 +5 \w^4 t +63 \w^2 t \de^2 \no\\&&\hspace{2cm}
   +36 \w^2 \de^4 +18 t \de^4) \Big\}\no\\
&+&J_0^\prime(-\de)\Big\{
\left(\frac{25 \gpnd^2 g_1^2}{729 \w } +\frac{34 \gpnd^4}{81 (\w^2-\de^2)}\w \right)(\de^2-M^2) (2 M^2 -2 \w^2 -t) \no\\&&\hspace{1cm}
+\frac{4 \gpnd^2}{9} \w (-M_\pi^2 +\de^2) \no\\&&\hspace{1cm}
-\frac{50 \gpnd^2 g_A g_1}{243 \w} (-2 M_\pi^4 +2 M_\pi^2 \w^2 + M_\pi^2 t +2 M_\pi^2 \de^2 -2 \w^2 \de^2 -t \de^2)\no\\&&\hspace{1cm}
+\frac{\gpnd^2 g_A^2}{3 \w} (-2 M_\pi^4 +2 \w^2 M_\pi^2 + M_\pi^2 t +2 M_\pi^2 \de^2 -2 \w^2 \de^2 -t \de^2) \Big\} \no\\
&-& K_0(t,-\de) \frac{4 \gpnd^2}{9} \de \w (-2 M_\pi^2 + t + 2 \de^2)\no\\
&-& I_0(t) \frac{2 \gpnd^2}{27} \w (-8 M_\pi^2 +12 \de^2 +5 t)\no\\
&-&\frac{\gpnd^2}{209952 \pi^2 (\w^2-\de^2)^2} \w ( 1944 M_\pi^4 (\w^2-\de^2) -5832 M_\pi^2 \w^2 \de^2 -972 M_\pi^2 \w^2 t \no\\&&\hspace{2cm}
   +1944 M_\pi^2 \w^4 +972 M_\pi^2 t \de^2 +3888 M_\pi^2 \de^4 -2268 t (\w^2 -\de^2)^2)\no\\&&
-\frac{\gpnd^2 g_1^2}{209952 \pi^2 (\w^2-\de^2)^2} \w ( -5400 \pi M_\pi^5 \de -3030 M_\pi^4 (\w^2-\de^2) \no\\&&\hspace{2cm}
   +2700 \pi M_\pi^3 \de (t +2 \w^2) -3790 M_\pi^2 \w^2 \de^2 +1260 M_\pi^2 \de^4 +2530 M_\pi^2 \w^4 \no\\&&\hspace{2cm}
   -1515 M_\pi^2 t \de^2 +1515 M_\pi^2 \w^2 t +500 \w^6 +760 \w^4 \de^2 -1260 \w^2 \de^4 \no\\&&\hspace{2cm}
   -630 t \de^4 +380 \w^2 t \de^2 +250 \w^4 t) \no\\
&&
-\frac{\gpnd^4}{209952 \pi^2 (\w^2-\de^2)^2} \w (16200 M_\pi^4 \de^2 -4536 M_\pi^4 \w^2 +2268 M_\pi^2 \w^2 t  \no\\&&\hspace{2cm}
   -5400 M_\pi^2 \w^2 \de^2 -18576 M_\pi^2 \de^4 -8100 M_\pi^2 t \de^2 +4536 M_\pi^2 \w^4 \no\\&&\hspace{2cm}
   -5400 \w^2 t \de^2 +9288 t \de^4 -10800 \w^4 \de^2 +18576 \w^2 \de^4)\no\\&&
+\frac{5 \gpnd^2 g_A g_1}{3888 \pi^2 (\w^2-\de^2)^2} \w ( 40 \pi M_\pi^5 \de -42 M_\pi^4 (\w^2-\de^2) -20 \pi M_\pi^3 \de (t+2 \w^2) \no\\&&\hspace{2cm}
   +M_\pi^2 (\w^2-\de^2)(21 t+14 \w^2 -14 \de^2) + 14 (\w^2-\de^2)^2 (2 \w^2 +t))\no\\&&
-\frac{\gpnd^2 g_A^2}{1296 \pi^2 \w \de (\w^2-\de^2)^2} (\pi M_\pi^5 (144 \de^4 +192w^4 -372 \w^2 \de^2) -96 M_\pi^4 \w^2 \de (\w^2 -\de^2) \no\\&&\hspace{2cm}
   +\pi M_\pi^3 (t+2 \w^2) (186 \w^2 \de^2 -96 \w^4 -72 \de^4) +48 M_\pi^2 \w^4 t \de -48 M_\pi^2 \w^2 t \de^3 \no\\&&\hspace{2cm}
   -106 M_\pi^2 \w^2 \de^5 +100 M_\pi^2 \w^4 \de^3 +6 M_\pi^2 \w^6 \de -98 \w^4 t \de^3 +53 \w^2 t \de^5 \no\\&&\hspace{2cm}
   +106 \w^4 \de^5 +90 \w^8 \de-196 \w^6 \de^3 +45 \w^6 \de t)~.\\
F_\pi^4 h^-(\w,t) & = & 
(J_0(\w)+J_0(-\w)) \frac{\w^2-M_\pi^2}{\w^2-\de^2} \Big\{
\frac{4 \gpnd^4}{243 (\w^2-\de^2)}(9 \de^2+8 \w^2) - \frac{8 \gpnd^2 g_A^2}{27} \Big\}\no\\
&+&(J_0(\w)-J_0(-\w)) \frac{\w^2-M_\pi^2}{\w^2-\de^2} \Big\{
-\frac{32 \gpnd^4}{243 (\w^2-\de^2)}\ w \de  + \frac{4 \gpnd^2 g_A^2}{27 \w} \de  \Big\}\no\\
&-& J_0(\de) \frac{2 \gpnd^4}{27 (\w^2-\de^2)^2} (M_\pi^2 - \de^2) (\w^2 +\de^2)\no\\
&+&J_0(-\de) \Big\{
\frac{\gpnd^4}{2187 (\w^2-\de^2)^2} (414 M_\pi^2 \w^2 +486 M_\pi^2 \de^2 +1710 \de^4 -2610 \w^2 \de^2)\no\\&&\hspace{1cm}
+\frac{ \gpnd^2}{9} 
-\frac{50 \gpnd^2 g_1^2}{2187 (\w^2-\de^2) \w^2} (\w^2+\de^2) (M_\pi^2 -\de^2) \no\\&&\hspace{1cm}
-\frac{100 \gpnd^2 g_A g_1}{243 \w^2} (M_\pi^2 -\de^2)
+\frac{2 \gpnd^2 g_A^2}{9 \w^2 (\w^2 -\de^2)} (M_\pi^2-\de^2) (w^2 -3\de^2) \Big\} \no\\
&+&\frac{1}{\w^2} \left(\frac{J_0(\w-\de)}{(\w-\de)^2} + \frac{J_0(-\w-\de)}{(\w+\de)^2}\right)\Big\{
\frac{25 \gpnd^2 g_1^2}{8748} (-5 M_\pi^2 \w^2 -4 M_\pi^2 \de^2 +5 \w^4  \no\\&&\hspace{2cm}
   +9 \w^2 \de^2 +4 \de^4)\no\\&&\hspace{1cm}
-\frac{25 \gpnd^2 g_A g_1}{486} ( -M_\pi^2 \w^2 -4 M_\pi^2 \de^2 + \w^4 +13 \w^2 \de^2 +4 \de^4)\no\\&&\hspace{1cm}
+\frac{\gpnd^2 g_A^2}{108} (-5 M_\pi^2 \w^2 -36 M_\pi^2 \de^2 +5 \w^4 +121 \w^2 \de^2 +36 \de^4)\Big\}\no\\
&+&\frac{\de}{\w} \left(\frac{J_0(\w-\de)}{(\w-\de)^2} - \frac{J_0(-\w-\de)}{(\w+\de)^2}\right)\Big\{
\frac{25 \gpnd^2 g_1^2}{4374} (-5 \w^2-4 \de^2)\no\\&&\hspace{1cm}
-\frac{25 \gpnd^2 g_A g_1}{243} ( 2 M_\pi^2 -3 \w^2 -6 \de^2)
+\frac{\gpnd^2 g_A^2}{54} (20 M_\pi^2 -25 \w^2 -56 \de^2)\Big\}\no\\
&-&J_0^\prime (-\de) \frac{122 \gpnd^4}{243 (\w^2-\de^2)} \de (M_\pi^2-\de^2)\no\\
&-& K_0(t,-\de) \frac{\gpnd^2}{18} (-4 M_\pi^2 +4 \de^2 +t)\no\\
&-& I_0(t) \frac{2 \gpnd^2}{9} \de \no\\
&-& \frac{\gpnd^4}{52488 \pi^2 (\w^2-\de^2)^2} \de ( 1458 M_\pi^2 (\w^2+\de^2) -108 \de^4 -2700  \de^2 \w^2 + 864 \w^4)\no\\&&
+\frac{\gpnd^2 g_1^2}{104976 \pi^2 (\w^2-\de^2)^2} ( 675 \pi M_\pi^3 (\w^2 + \de^2) + 645 M_\pi^2 \de (\w^2-\de^2) +4850 \de \w^4  \no\\&&\hspace{2cm}
   +5010 \de^5 -9860 \w^2 \de^3)\no\\&&
+\frac{\gpnd^2}{108 \pi^2 (\w^2-\de^2)}\de (M_\pi^2 -5 \w^2 +5 \de^2)\no\\&&
-\frac{5 \gpnd^2 g_A g_1}{1944 \pi^2 (\w^2-\de^2)^2} ( \pi M_\pi^3 (15\w^2 -25 \de^2) +3 M_\pi^2 \de (\w^2-\de^2) +70 \de (\w^2-\de^2)^2)\no\\&&
+\frac{\gpnd^2 g_A^2}{1296 \pi^2 (\w^2-\de^2)^2} (\pi M_\pi^3 (117 \w^2 -123 \de^2)+72 M_\pi^2 \de (\w^2-\de^2) +210 \de^5 \no\\&&\hspace{2cm}
   -388 \w^2 \de^3 +178 \w^4 \de)~.
\end{eqnarray}

\section{Threshold parameters}
\setcounter{equation}{0}
\label{app:sub}

Here, we collect the analytical expressions of the $\Delta$--contribution for the S-- and P--wave
scattering lengths and effective ranges. The corresponding pure
nucleon terms are given in Ref.\cite{FM}. As before, we explicitly
display the off--shell parameters but we note that these are fixed in
the actual calculation. We have:
\begin{eqnarray}
a_{0+}^- & = & \frac{\gpnd^2 M_\pi^3}{36 \pi (m+M_\pi) m F^2} (1-4 z_0 +4 z_0^2) -\frac{M_\pi^3 (b_3+b_8)}{18 \pi (m+M_\pi) F^2} (-5 +4 z +4 z_0 +4 z z_0)~,\\
b_{0+}^- & = & \frac {\gpnd^2 M_\pi}{144\pi (m+M_\pi) m^3 (M_\pi+\de) F^2} \Big\{
16 m^3 + m^2 [ M_\pi^2 ( 2 -20 z_0 -24 z_0^2) -\de ( 30 +40 z_0 +24 z_0^2) ]\no\\&&\hspace{1cm}
- M_\pi (M_\pi+\de) (2 m - M_\pi) (1-4 z_0 +4 z_0^2) \Big\} \no\\&&
-\frac{\gpnd M_\pi (b_3+b_8)}{72 \pi (m+M_\pi) m^2 (M_\pi+\de) F^2} \Big\{
m^2 [ M_\pi(-26 -2 z+24 z_0+16 z z_0)\no\\&&\hspace{2cm}
+\de (-42-2 z+24 z_0+16 z z_0)] +M_\pi  (M_\pi + \de)(2 m - M_\pi)(5 -4 z -4 z_0 -4 z z_0) \Big\}\no\\&&
-\frac{\gpnd M_\pi b_8}{36 \pi (m+M_\pi) F^2} (6+13 z +14 z z_0) \no\\
&-&\frac{\gpnd^2 m M_\pi}{648 \pi^3 (m+M_\pi) \sqrt{\de^2-M_\pi^2} F^4} \Big\{
15 \de \ln{\frac{\de +\sqrt{\de^2-M_\pi^2}}{M_\pi}} + 13 \sqrt{\de^2-M_\pi^2} \Big\} \\
a_{0+}^+ & = & \frac{\gpnd^2 M_\pi^2}{18 \pi (m+M_\pi) m F^2} \Big\{
m (-5+8 z_0 +4 z_0^2) +6 \de(1-2 z_0)\Big\}\no\\
&-&\frac{\gpnd^2 m M_\pi^2 \sqrt{\de^2-M_\pi^2}}{24 \pi^3 (m+M_\pi) F^4} \ln{\frac{\de +\sqrt{\de^2-M_\pi^2}}{M_\pi}}~,\\
b_{0+}^+ & = & \frac{\gpnd^2}{72 \pi (m+M_\pi) m^2 (M_\pi+\de) F^2} \Big\{
4 m^3 [ M_\pi(-1 +8 z_0 +4 z_0^2) +\de (-5 +8 z_0 +4 z_0^2) ]\no\\&&\hspace{1cm}
+2 m^2 [ M_\pi^2 (9 -8 z_0 -20 z_0^2) + M_\pi \de ( 5 -32 z_0 -20 z_0^2) +12 \de^2 (1-2 z_0)]\no\\&&\hspace{1cm}
+m [ M_\pi^3 (-5 +8 z_0 +4 z_0^2)+ M_\pi^2 \de (-17 +32 z_0 +4 z_0^2) + 12 M_\pi \de^2 (-1 +2 z_0)]\no\\&&\hspace{1cm}
+ 6 M_\pi^2 (M_\pi+\de) \de (1-2 z_0) \Big\} \no\\&&
-\frac{4 \gpnd (b_3 + b_8) M_\pi^2}{9 \pi (m+M_\pi) ( M_\pi+\de) F^2}\no\\
&-&\frac{\gpnd^2}{864 \pi^3 (m+M_\pi) m \sqrt{\de^2-M_\pi^2} F^4} \ln{\frac{\de +\sqrt{\de^2-M_\pi^2}}{M_\pi}} \Big\{
m^2 (-154 M_\pi^2 +144\de^2) \no\\&&\hspace{1cm}
+ 9 M_\pi (2 m-M_\pi) (M_\pi^2- \de^2) \Big\}~,\\
a_{1-}^- & = & \frac{\gpnd^2}{54 \pi (m+M_\pi) m (M_\pi+\de) F^2} \Big\{
2 m^2 + m [M_\pi(5 +8 z_0 +12 z_0^2) +\de (1+8 z_0 +12 z_0^2)]\no\\&&\hspace{1cm}
+M_\pi^2 (5 +12 z_0 +16 z_0^2) -3 M_\pi \de -2 \de^2(1+6 z_0 +8 z_0^2)\Big\}\no\\&&
+\frac{2 \gpnd M_\pi (b_3+b_8)}{27 \pi (m+M_\pi) (M_\pi+\de) F^2} \Big\{ -m + M_\pi (-2 +z_0) +\de z_0\Big\}\no\\&&
+\frac{\gpnd M_\pi b_8}{54\pi (m+M_\pi) F^2} (1+4 z +12 z z_0)\no\\&&
- \frac{m [2\gpnd (e_1+e_2)-4 \gpnd (2 e_4-e_5)-(b_3+b_8)^3]}{27\pi (m+M_\pi) (M_\pi+\de) F^2}\Big\}\no\\
&
+&\frac{\gpnd^2 m}{3888 \pi^3 (m+M_\pi) (\de^2-M_\pi^2)^2 \sqrt{\de^2-M_\pi^2} M_\pi^2 F^4} \no\\
&&
\Big\{\sqrt{\de^2-M_\pi^2} \sqrt{\de(2 M_\pi- \de)} \left( \arcsin{\frac{M_\pi-\de}{M_\pi}} +\frac{\pi}{2}\right) \no\\&&\hspace{1cm}
\Big[\frac{125 g_1^2}{27}(\de-M_\pi) \de (-2 M_\pi^4-M_\pi^3 \de +3 M_\pi^2 \de^2+M_\pi \de^3 - \de^4)\no\\&&\hspace{1cm}
-\frac{50 g_A g_1}{3} (\de-M_\pi) \de (-2 M_\pi^4 +3 M_\pi^3 \de +9 M_\pi^2 \de^2 + M_\pi \de^3 -3 \de^4)\no\\&&\hspace{1cm}
-\frac{g_A^2}{2}\de (-60 M_\pi^5 + 270 M_\pi^4 \de+216 M_\pi^3 \de^2 -456 M_\pi^2 \de^3-180 M_\pi \de^4 +162 \de^5) \Big]\no\\
&&
+\sqrt{\de^2-M_\pi^2} \Big[
-(\de^2-M_\pi^2) M_\pi^2 (29 M_\pi^3 -24 M_\pi^2 \de -26 M_\pi \de^2 +30 \de^3)\no\\&&\hspace{1cm}
+\frac{5 g_1^2}{324} (\de-M_\pi) M_\pi [ ( 533- 270 \pi) M_\pi^5 + (-1485 + 270 \pi) M_\pi^4 \de -2702 M_\pi^3 \de^2 \no\\&&\hspace{1cm}
   +1410 M_\pi^2 \de^3 +2454 M_\pi \de^4 + 360 \de^5 ]
+\frac{\gpnd^2}{3} M_\pi^2 (-123 M_\pi^5 -28 M_\pi^4 \de +221 M_\pi^3 \de^2 \no\\&&\hspace{1cm}
   + 38 M_\pi^2 \de^3 -80 M_\pi \de^4 -46 \de^5)
-\frac{5 g_A g_1}{6} (\de-M_\pi) M_\pi [ (55 -30 \pi) M_\pi^5 \no\\&&\hspace{1cm}
   -(31 +50 \pi) M_\pi^4 \de -170 M_\pi^3 \de^2 +6 M_\pi^2 \de^3 +130 M_\pi \de^4 +40  \de^5]\no\\&&\hspace{1cm}
-\frac{g_A^2}{2 \de} M_\pi [-96 \pi M_\pi^7 + (243 - 117 \pi)M_\pi^6 \de - (184 -186 \pi) M_\pi^5 \de^2 \no\\&&\hspace{1cm}
   - (545 - 123 \pi) M_\pi^4 \de^3 + (331-72\pi) M_\pi^3 \de^4 + 410 M_\pi^2 \de^5 - 147 M_\pi \de^6 -108 \de^7 ]\Big] \no\\
&&
+\sqrt{\de^2-M_\pi^2} \sqrt{\de(\de+2 M_\pi)} \ln{\frac{\de+M_\pi+\sqrt{\de(\de+2M_\pi)}}{M_\pi}} \no\\&&\hspace{1cm}
\Big[
\frac{100 g_1^2}{108} (\de-M_\pi) \de^2 (20 M_\pi^3 -8 M_\pi^2 \de -11 M_\pi \de^2 -\de^3)\no\\&&\hspace{1cm}
-\frac{50 g_A g_1}{3} (\de-M_\pi) \de^3 (-2 M_\pi^2 + M_\pi \de +\de^2)\no\\&&\hspace{1cm}
-3 g_A^2 \de^2 (-20 M_\pi^4 + 12 M_\pi^3 \de +27 M_\pi^2 \de^2- 10 M_\pi \de^3 -9 \de^4)\Big]\no\\
&&
+\ln{\frac{\de+\sqrt{\de^2-M_\pi^2}}{M_\pi}} \Big[
-6 (\de^2-M_\pi^2) M_\pi^2 (-6 M_\pi^4 +5 M_\pi^3 \de +12 M_\pi^2 \de^2 -5 M_\pi \de^3 -6 \de^4)\no\\&&\hspace{1cm}
+\frac{100 g_1^2}{108} (\de-M_\pi) (-4 M_\pi^7 + 22 M_\pi^6 \de + 30 M_\pi^5 \de^2 -48 M_\pi^4 \de^3 \no\\&&\hspace{1cm}
   -48 M_\pi^3 \de^4 + 30 M_\pi^2 \de^5 + 22 M_\pi \de^6 -4 \de^7)
+\frac{2 \gpnd^2}{3} M_\pi^2 (-64 M_\pi^6 -210 M_\pi^5 \de \no\\&&\hspace{1cm}
   + 422 M_\pi^4 \de^2 + 420 M_\pi^3 \de^3 -652 M_\pi^2 \de^4 -210 M_\pi \de^5 +294 \de^6)\no\\&&\hspace{1cm}
-\frac{100 g_A g_1}{3} (\de-M_\pi) (2 M_\pi^7 +5 M_\pi^6 \de -3 M_\pi^5 \de^2 -12 M_\pi^4 \de^3 + 9 M_\pi^2 \de^5 + M_\pi \de^6 -2 \de^7)\no\\&&\hspace{1cm}
-\frac{6 g_A^2}{\de}(16 M_\pi^9 +6 M_\pi^8 \de - 29 M_\pi^7 \de^2 -36 M_\pi^6 \de^3 -9 M_\pi^5 \de^4 +72 M_\pi^4 \de^5 +41 M_\pi^3 \de^6 \no\\&&\hspace{1cm}
   -60 M_\pi^2 \de^7 -19 M_\pi \de^8 +18 \de^9)\Big]
\Big\}~, \\
a_{1-}^+ & = & -\frac{\gpnd^2}{54 \pi (m+M_\pi) m (M_\pi+\de) F^2} \Big\{
-4 m^2 + m [ M_\pi (-7 +8 z_0 + 12 z_0^2) + \de (1 +8 z_0 +12 z_0^2) ]\no\\&&\hspace{1cm}
+ 4 M_\pi^2 (-1+4 z_0^2) +6 M_\pi \de (1-2 z_0)-2 \de^2 (1+6 z_0 +8 z_0^2) \Big\} \no\\&&
-\frac{4 \gpnd M_\pi (b_3+b_8)}{27 \pi (m+M_\pi) (M_\pi+\de) F^2} (m+2 M_\pi (1+z_0) +2 \de z_0)\no\\&&
-\frac{2 \gpnd M_\pi b_8}{27 \pi (m+M_\pi) F^2} (1+4 z +12 z z_0)\no\\&&
-\frac{2 M_\pi^2 m [ 2\gpnd (e_1+e_2) -4\gpnd (2 e_4-e_5) - (b_3+b_8)^2]}{27 \pi (m+M_\pi) ( M_\pi+\de) F^2}\no\\
&+&\frac{\gpnd^2 m}{324 \pi^3 (m+M_\pi) (\de^2-M_\pi^2)^2 \sqrt{\de^2-M_\pi^2} M_\pi^2 F^4} \no\\&&
\Big\{
\ln{\frac{\de+\sqrt{\de^2-M_\pi^2}}{M_\pi}} \Big[
\frac{1}{4} M_\pi^2 (\de^2-M_\pi^2) (77 M_\pi^4 -149 M_\pi^2 \de^2 +72 \de^4)\no\\&&\hspace{1cm}
-\frac{25 g_1^2}{162} (\de-M_\pi)(- M_\pi^7 -27 M_\pi^6 \de - 15 M_\pi^5 \de^2 + 63 M_\pi^4 \de^3 + 33 M_\pi^3 \de^4 -45 M_\pi^2 \de^5 \no\\&&\hspace{1cm}
   -17 M_\pi \de^6 +9 \de^7)
-\frac{2 \gpnd^2}{9} M_\pi^2 (16 M_\pi^6 + 105 M_\pi^5 \de -147 M_\pi^4 \de^2 -210 M_\pi^3 \de^3 \no\\&&\hspace{1cm}
   +246 M_\pi^2 \de^4 +105 M_\pi \de^5 -115 \de^6)
+\frac{25 g_A g_1}{9} (\de-M_\pi) (-M_\pi^7 -7 M_\pi^6 \de -3 M_\pi^5 \de^2 \no\\&&\hspace{1cm}
   +15 M_\pi^4 \de^3 +9 M_\pi^3 \de^4 -9 M_\pi^2 \de^5 -5 M_\pi \de^6 + \de^7)
+\frac{g_A^2}{2 \de} (-32 M_\pi^9 - M_\pi^8 \de + 58 M_\pi^7 \de^2 \no\\&&\hspace{1cm}
   +12 M_\pi^6 \de^3 +18 M_\pi^5 \de^4 -30 M_\pi^4 \de^5 - 82 M_\pi^3 \de^6 +28 M_\pi^2 \de^7 + 38 M_\pi \de^8 -9 \de^9) 
\Big]\no\\
&&
+\sqrt{\de^2-M_\pi^2} \Big[
\frac{1}{2} M_\pi^4 (\de^2-M_\pi^2) (2 M_\pi +\de)
-\frac{5 g_1^2}{1944} (\de-M_\pi) M_\pi [(-1118+270 \pi)M_\pi^5 \no\\&&\hspace{1cm}
   -(285+270 \pi) M_\pi^4 \de + 2387 M_\pi^3 \de^2 + 930 M_\pi^2 \de^3 -984 M_\pi \de^4  -360 \de^5]\no\\&&\hspace{1cm}
-\frac{\gpnd^2}{9} M_\pi^2 (51 M_\pi^5 -26 M_\pi^4 \de -102 M_\pi^3 \de^2 +44 M_\pi^2 \de^3 + 33 M_\pi \de^4 -9 \de^5)\no\\&&\hspace{1cm}
+\frac{5 g_A g_1}{36} (\de-M_\pi) M_\pi [ -(50+10 \pi) M_\pi^5 - (23-10 \pi) M_\pi^4 \de +121 M_\pi^3 \de^2 +78 M_\pi^2 \de^3 \no\\&&\hspace{1cm}
    -56 M_\pi \de^4 -40 \de^5]
+\frac{g_A^2}{12 \de} M_\pi [96\pi M_\pi^7-(222-21 \pi) M_\pi^6 \de + (1-186\pi) M_\pi^5 \de^2 \no\\&&\hspace{1cm}
   + (491-75\pi) M_\pi^4 \de^3 -(25 - 77 \pi) M_\pi^3 \de^4 -377 M_\pi^2 \de^5+ 24 M_\pi \de^6+108 \de^7] \Big]\no\\
&& 
+\sqrt{\de^2-M_\pi^2} \sqrt{\de(2 M_\pi- \de)} \left( \arcsin{\frac{M_\pi-\de}{M_\pi}} +\frac{\pi}{2}\right) \no\\&&\hspace{1cm}
\Big[-\frac{125 g_1^2}{108} (\de-M_\pi) \de (2 M_\pi^4 + M_\pi^3 \de -3 M_\pi^2 \de^2 - M_\pi \de^3 + \de^4)\no\\&&\hspace{1cm}
+\frac{25 g_A g_1}{6} (\de-M_\pi) \de (2 M_\pi^4 + M_\pi^3 \de -3 M_\pi^2 \de^2 - M_\pi \de^3 + \de^4)\no\\&&\hspace{1cm}
+\frac{3 g_A^2}{8}\de (20 M_\pi^5 -10 M_\pi^4 \de -24 M_\pi^3 \de^2 +44 M_\pi^2 \de^3 +20 M_\pi \de^4 -18 \de^5) \Big]\no\\
&&
+\sqrt{\de^2-M_\pi^2} \sqrt{\de(\de+2 M_\pi)} \ln{\frac{\de+M_\pi+\sqrt{\de(\de+2M_\pi)}}{M_\pi}}  \no\\&&\hspace{1cm}
\Big[-\frac{25 g_1^2}{324} (\de-M_\pi) \de (10 M_\pi^4 -25 M_\pi^3 \de +11 M_\pi^2 \de^2 +7 M_\pi \de^3 -3 \de^4) \no\\&&\hspace{1cm}
+\frac{25 g_A g_1}{18} (\de-M_\pi)\de (2 M_\pi^4 -5 M_\pi^3 \de - M_\pi^2 \de^2 +3 M_\pi \de^3 + \de^4) \no\\&&\hspace{1cm}
+\frac{g_A^2}{4}\de (10 M_\pi^5 -35 M_\pi^4 \de +12 M_\pi^3 \de^2 + 32
M_\pi^2 \de^3 -10 M_\pi \de^4 -9 \de^5) \Big] \Big\} ~,\\
a_{1+}^- & = & -\frac{\gpnd^2}{108 \pi (m+M_\pi) m (\de^2-M_\pi^2) F^2} \Big\{
m^2 (-4 M_\pi -2 \de) + m[ M_\pi^2 (-1 + 8 z_0 +12 z_0^2) \no\\&&\hspace{1cm}
   + 2 M_\pi \de -\de^2(1+8 z_0 +12 z_0^2) ]
+ M_\pi^3 (-1 +4 z_0^2) + 4 M_\pi^2\de (1 -3 z_0 -4 z_0^2)\no\\&&\hspace{1cm}
   - M_\pi \de^2 (5 +4 z_0^2) +2 \de^3 (1 +6 z_0 +8 z_0^2) \Big\}\no\\&&
+\frac{\gpnd M_\pi (b_3+b_8)}{27 \pi (m+M_\pi) (\de^2-M_\pi^2) F^2}\Big\{
m (M_\pi + 2 \de)+ M_\pi^2 (2 z_0-1) + M_\pi \de -2 \de^2 z_0\Big\}\no\\&&
+\frac{\gpnd M_\pi b_8}{54 \pi (m+M_\pi) F^2} (1+4 z +12 z z_0)\no\\&&
+\frac{M_\pi^2 m [ 2\gpnd (e_1+e_2) -4 \gpnd (2e_4-e_5) - (b_3+b_8)^2 ]}{54 \pi (m+M_\pi) (\de^2-M_\pi^2) F^2} (2 M_\pi + \de) \no\\
&+&\frac{\gpnd^2 m}{3888 \pi^3 (m+M_\pi) (\de^2-M_\pi^2)^2 \sqrt{\de^2-M_\pi^2} M_\pi^2 F^4} \no\\&&
\Big\{\sqrt{\de^2-M_\pi^2} 
\Big[M_\pi^2 (\de^2-M_\pi^2) (\de-M_\pi) (29 M_\pi^2 + 41 M_\pi \de + 15 \de^2)\no\\&&\hspace{1cm}
-\frac{5 g_1^2}{324} M_\pi [ (533+136\pi) M_\pi^6+(1009+540\pi) M_\pi^5\de -(1217-135\pi) M_\pi^4\de^2\no\\&&\hspace{1cm}
   -2056 M_\pi^3 \de^3+ 1044 M_\pi^2 \de^4 + 1047 M_\pi \de^5-360 \de^6]\no\\&&\hspace{1cm}
+\frac{\gpnd^2}{3} (M_\pi+\de) M_\pi^2 (-123 M_\pi^4 + 137 M_\pi^3 \de + 84 M_\pi^2 \de^2 -103 M_\pi \de^3 + 23 \de^4)\no\\&&\hspace{1cm}
-\frac{5 g_A g_1}{6} M_\pi (-(55+15 \pi) M_\pi^6- (43-20 \pi) M_\pi^5 \de + (139+25\pi) M_\pi^4 \de^2 + 88 M_\pi^3 \de^3\no\\&&\hspace{1cm}
   -124 M_\pi^2 \de^4 -45 M_\pi \de^5+ 40 \de^6)
-\frac{g_A^2}{4 \de} M_\pi (-192 \pi M_\pi^7 + ( 486 + 117\pi ) M_\pi^6 \de \no\\&&\hspace{1cm}
   + ( 184 + 372 \pi ) M_\pi^5 \de^2- ( 1090 + 123 \pi ) M_\pi^4 \de^3 -(331+144 \pi) M_\pi^3 \de^4+ 820 M_\pi^2 \de^5 \no\\&&\hspace{1cm}
   + 147 M_\pi \de^6- 216 \de^7) \Big] \no\\
&&
+\ln{\frac{\de+\sqrt{\de^2-M_\pi^2}}{M_\pi}} \Big[
6 M_\pi^2 (\de^2-M_\pi^2) (\de-M_\pi) (3 M_\pi^3+8 M_\pi^2\de +2 M_\pi \de^2 -3 \de^3)\no\\&&\hspace{1cm}
-\frac{50 g_1^2}{27} (M_\pi^8 +13 M_\pi^7 \de -2 M_\pi^6 \de^2 -39 M_\pi^5 \de^3 + 39 M_\pi^3 \de^5+2 M_\pi^2 \de^6 -13 M_\pi \de^7 - \de^8)\no\\&&\hspace{1cm}
+\frac{2 \gpnd^2}{3} (M_\pi+\de) M_\pi^2 (32 M_\pi^5  -242 M_\pi^4 \de + 31 M_\pi^3 \de^2 + 389 M_\pi^2 \de^3 -63 M_\pi \de^4 -147 \de^5)\no\\&&\hspace{1cm}
-\frac{100 g_A g_1}{3} ( M_\pi^8 -3 M_\pi^7 \de -4 M_\pi^6 \de^2 +9 M_\pi^5 \de^3 + 6 M_\pi^4 \de^4 -9 M_\pi^3 \de^5 -4 M_\pi^2 \de^6\no\\&&\hspace{1cm}
   +3 M_\pi \de^7+ \de^8)
-\frac{6 g_A^2}{\de} (16 M_\pi^9 -3 M_\pi^8 \de-29 M_\pi^7 \de^2 + 18 M_\pi^6 \de^3 -9 M_\pi^5 \de^4 \no\\&&\hspace{1cm}
   -36 M_\pi^4 \de^5 +41 M_\pi^3 \de^6 + 30 M_\pi^2 \de^7 -19 M_\pi \de^8 -9 \de^9) \Big]\no\\
&&
+\sqrt{\de^2-M_\pi^2} \sqrt{\de(2 M_\pi-\de)} \left( \arcsin{\frac{M_\pi-\de}{M_\pi}} +\frac{\pi}{2}\right) \no\\&&\hspace{1cm}
\Big[
-\frac{25 g_1^2}{108} \de (-10 M_\pi^5 + 65 M_\pi^4 \de + 104 M_\pi^3 \de^2 -19 M_\pi^2 \de^3 -40 M_\pi \de^4 + 8 \de^5)\no\\&&\hspace{1cm}
-\frac{25 g_A g_1}{6}  M_\pi \de (2 M_\pi^4 -5 M_\pi^3 \de -12 M_\pi^2 \de^2 - M_\pi \de^3 +4 \de^4)\no\\&&\hspace{1cm}
-\frac{15 g_A^2}{4} M_\pi^2 \de (-2 M_\pi^3 -3 M_\pi^2 \de + \de^3) \Big] \no\\
&&
+\sqrt{\de^2-M_\pi^2} \sqrt{\de(\de+2 M_\pi)} \ln{\frac{\de+M_\pi+\sqrt{\de(\de+2M_\pi)}}{M_\pi}} \no\\&&\hspace{1cm}
\Big[-\frac{25 g_1^2}{108} \de( 30 M_\pi^5 +35 M_\pi^4 \de -88 M_\pi^3 \de^2 -33 M_\pi^2 \de^3+40 M_\pi \de^4 + 16 \de^5)\no\\&&\hspace{1cm}
-\frac{25 g_A g_1}{6}  \de ( -6 M_\pi^5 -15 M_\pi^4 \de + 20 M_\pi^3 \de^2 +21 M_\pi^2 \de^3 -12 M_\pi \de^4 -8 \de^5)\no\\&&\hspace{1cm}
-\frac{3 g_A^2}{4} \de (30 M_\pi^5 +115 M_\pi^4 \de -96 M_\pi^3 \de^2 -201 M_\pi^2 \de^3 +80 M_\pi \de^4 + 72 \de^5)\Big]
\Big\} ~,\\
a_{1+}^+ & = & \frac{\gpnd^2}{54 \pi (m+M_\pi) m (\de^2-M_\pi^2) F^2} \Big\{
2 m^2 (M_\pi+2 \de) + m [M_\pi^2 (-1 +8 z_0 +12 z_0^2)\no\\&&\hspace{1cm}
   +2 M_\pi \de - \de^2 (1+8 z_0 +12 z_0^2)] + M_\pi^3 (-1 +4 z_0^2) +4 M_\pi^2 \de (1-3 z_0 -4 z_0^2) \no\\&&\hspace{1cm}
   - M_\pi \de^2 (5+4 z_0^2) +2 \de^3 (1+6 z_0 +8 z_0^2) \Big\}\no\\&&
+\frac{2 \gpnd M_\pi (b_3 +b_8)}{27 \pi (m+M_\pi) (\de^2-M_\pi^2) F^2} \Big\{
m (2 M_\pi+\de) + M_\pi^2 (1-2 z_0) - M_\pi \de +2 \de^2 z_0 \Big\}\no\\&&
+\frac{\gpnd M_\pi b_8}{27 \pi (m+M_\pi) F^2} (1+4 z +12 z z_0)\no\\&&
-\frac{M_\pi^2 m (M_\pi+2 \de) [ 2\gpnd (e_1+e_2) -2 \gpnd (2 e_4 - e_5) -(b_3-b_8)^2]}{27 \pi (m+M_\pi) (\de^2-M_\pi^2) F^2} \no\\
&+&\frac{\gpnd^2 m}{648 \pi^3 (m+M_\pi) (\de^2-M_\pi^2)^2 \sqrt{\de^2-M_\pi^2} M_\pi^2 F^4} \no\\&&
\Big\{\sqrt{\de^2-M_\pi^2} \Big[
M_\pi^4 (\de-M_\pi) (\de^2-M_\pi^2) 
+\frac{5 g_1^2}{972} M_\pi [ (559 + 270 \pi) M_\pi^6+ ( 833+ 270 \pi) M_\pi^5 \de \no\\&&\hspace{1cm}
   -(1336-270\pi) M_\pi^4 \de^2-1457 M_\pi^3 \de^3+ 957 M_\pi^2 \de^4+ 624 M_\pi \de^5-180 \de^6]\no\\&&\hspace{1cm}
+\frac{\gpnd^2}{9} (\de+M_\pi) M_\pi^2 (51 M_\pi^4 + M_\pi^3 \de -103 M_\pi^2 \de^2 +15 M_\pi \de^3 + 18 \de^4] \no\\&&\hspace{1cm}
-\frac{5 g_A g_1}{18} M_\pi [(25-10 \pi) M_\pi^6+ (27 -10 \pi ) M_\pi^5 \de - (77+10 \pi ) M_\pi^4 \de^2-43 M_\pi^3 \de^3 \no\\&&\hspace{1cm}
   + 67 M_\pi^2 \de^4 + 16 M_\pi \de^5 -20 \de^6]
+\frac{g_A^2}{12\de}M_\pi [-96\pi M_\pi^7 +(222+42 \pi) M_\pi^6 \de \no\\&&\hspace{1cm}
   + (2+186\pi) M_\pi^5 \de^2 - (491+150 \pi ) M_\pi^4 \de^3-(50+72\pi) M_\pi^3 \de^4+377 M_\pi^2 \de^5+ 48 M_\pi \de^6\no\\&&\hspace{1cm}
   -108 \de^7]\Big] \no\\
&&
+\ln{\frac{\de+\sqrt{\de^2-M_\pi^2}}{M_\pi}} \Big[
-\frac{1}{2} (\de^2-M_\pi^2) (\de-M_\pi) \de M_\pi^2 (77 M_\pi^2 -72 \de^2)
+\frac{25 g_1^2}{81} (-M_\pi^8 + 13 M_\pi^7\de \no\\&&\hspace{1cm}
   +12 M_\pi^6 \de^2 -39 M_\pi^5 \de^3 -30 M_\pi^4 \de^4 +39 M_\pi^3 \de^5 +28 M_\pi^2 \de^6 -13 M_\pi \de^7 -9 \de^8)\no\\&&\hspace{1cm}
+\frac{2 \gpnd^2}{9} (\de+M_\pi) M_\pi^2 (-32 M_\pi^5 + 137 M_\pi^4 \de + 157 M_\pi^3 \de^2 -367 M_\pi^2 \de^3 -125 M_\pi \de^4 \no\\&&\hspace{1cm}
   + 230 \de^5)
-\frac{50g_A g_1}{9}(-M_\pi^8 +3 M_\pi^7 \de +4 M_\pi^6 \de^2 -9 M_\pi^5 \de^3 -6 M_\pi^4\de^4 +9 M_\pi^3 \de^5 \no\\&&\hspace{1cm}
   +4 M_\pi^2 \de^6-3 M_\pi \de^7 -\de^8)
+\frac{g_A^2}{\de} (16 M_\pi^9-M_\pi^8 \de-29 M_\pi^7 \de^2+12 M_\pi^6 \de^3-9 M_\pi^5 \de^4 \no\\&&\hspace{1cm}
   -30 M_\pi^4 \de^5+41 M_\pi^3 \de^6+28 M_\pi^2 \de^7-19 M_\pi \de^8 -9 \de^9) \Big]\no\\
&&
+\sqrt{\de^2-M_\pi^2} \sqrt{\de(2 M_\pi-\de)} \left( \arcsin{\frac{M_\pi-\de}{M_\pi}} +\frac{\pi}{2}\right) \no\\&&\hspace{1cm}
\Big[
\frac{25g_1^2}{324} \de (30 M_\pi^5 + 45 M_\pi^4 \de + 24 M_\pi^3 \de^2 + 21 M_\pi^2 \de^3 -12 \de^5)\no\\&&\hspace{1cm}
-\frac{25 g_A g_1}{6} M_\pi^2 \de (2 M_\pi^3 +3 M_\pi^2 \de - \de^3)
+\frac{15 g_A^2}{4}  M_\pi^2 \de(2 M_\pi^3+3 M_\pi^2 \de - \de^3) \Big]\no\\
&&
+\sqrt{\de^2-M_\pi^2} \sqrt{\de(\de+2 M_\pi)} \ln{\frac{\de+M_\pi+\sqrt{\de(\de+2M_\pi)}}{M_\pi}} \no\\&&\hspace{1cm}
\Big[
\frac{25 g_1^2}{324} \de (50 M_\pi^5 +5 M_\pi^4 \de -72 M_\pi^3 \de^2 -47 M_\pi^2 \de^3 +40 M_\pi \de^4 + 24 \de^5)\no\\&&\hspace{1cm}
-\frac{25g_A g_1}{18}\de (10 M_\pi^5 + M_\pi^4 \de -16 M_\pi^3 \de^2 -7 M_\pi^2 \de^3 +8 M_\pi \de^4 + 4 \de^5)\no\\&&\hspace{1cm}
+\frac{g_A^2}{4} \de (50 M_\pi^5 + 5 M_\pi^4 \de -48 M_\pi^3 \de^2 -83 M_\pi^2 \de^3 + 40 M_\pi \de^4 + 36 \de^5)\Big]
\Big\} ~.
\end{eqnarray}
We note that some of these terms are clearly of fourth order in the
chiral expansion, because in that case one counts $\Delta$ as order
${\cal O}(p^0)$.

\newpage

\newpage

\noindent {\Large {\bf Figures}}

$\,$

\vskip 4cm

\newpage

\vskip 1cm

\newpage

\vskip 1cm

\newpage

\vskip 1cm

\newpage

\vskip 1cm

\begin{thebibliography}{99}
\baselineskip 12pt
\bibitem{FMS} N. Fettes, Ulf-G. Mei{\ss}ner and S. Steininger, 
Nucl. Phys. A640 (1998) 199.\vs
\bibitem{FM}N. Fettes and Ulf-G. Mei{\ss}ner, {\tt hep-ph/0002162},
  Nucl. Phys. A (2000), in print.\vs
\bibitem{JMdel}E. Jenkins and A.V. Manohar, Phys. Lett. B259 (1991) 353.\vs
\bibitem{HHK}T.R. Hemmert, B.R. Holstein and J. Kambor, J. Phys. G:
  Nucl. Part. Phys.  24 (1998) 1831.\vs
\bibitem{bkmlec} V. Bernard, N. Kaiser and Ulf-G. Mei{\ss}ner,
Nucl. Phys. A615 (1997) 483.\vs
\bibitem{BL} T. Becher and H. Leutwyler, Eur. Phys. J.C9 (1999) 643.\vs
\bibitem{ET}P.J. Ellis and H.-B. Tang, Phys. Rev. C57  (1998) 3356.\vs
\bibitem{GaZe}J. Gasser and A. Zepeda, Nucl. Phys. B174 (1980) 445.\vs
\bibitem{JM}E. Jenkins and A.V. Manohar, Phys. Lett. B255 (1991) 558.\vs
\bibitem{BKKM}V. Bernard, N. Kaiser, J. Kambor and Ulf-G. Mei{\ss}ner,
Nucl. Phys. B388 (1992) 315.\vs
\bibitem{BKMrev} V. Bernard, N. Kaiser and Ulf-G. Mei{\ss}ner,
Int. J. Mod. Phys. E4 (1995) 193.\vs
\bibitem{BFHM}V. Bernard, H.W. Fearing, T.R. Hemmert and Ulf-G. Mei{\ss}ner,
Nucl. Phys. A635 (1998) 121; (E) Nucl. Phys. A642 (1998) 563.\vs
\bibitem{BKMres} V. Bernard, N. Kaiser and Ulf-G. Mei{\ss}ner,
Nucl. Phys. B364 (1991) 283.\vs
\bibitem{koch} R. Koch,  Nucl. Phys. A448 (1986) 707.\vs
\bibitem{mats} E. Matsinos, Phys. Rev. C56 (1997) 3014;
E. Matsinos, private communication.\vs
\bibitem{SAID}SAID on-line program,
R.A. Arndt et al., see website http://said.phys.vt.edu/.\vs
\bibitem{BHM} V. Bernard, T.R. Hemmert and Ulf-G. Mei{\ss}ner, {\it forthcoming}.\vs
\bibitem{HHKK}T.R. Hemmert, B.R. Holstein, J. Kambor and G. Kn\"ochlein,
Phys. Rev.  D57 (1998) 5746.\vs
\bibitem{bkmpin2} V. Bernard, N. Kaiser and Ulf-G. Mei{\ss}ner,
Phys. Rev. C52 (1995) 2185.\vs
\bibitem{DP}A. Datta and S. Pakvasa, Phys. Rev. D56  (1997) 4322.\vs
\bibitem{moj} M. Moj\v zi\v s,  Eur. Phys. J. C2  (1998) 181.\vs
\bibitem{gls} J. Gasser, H. Leutwyler and M.E. Sainio,
Phys. Lett. B253 (1990) 252.\vs
\bibitem{bkmcd} V. Bernard, N. Kaiser and Ulf-G. Mei{\ss}ner,
Phys. Lett. B389 (1996) 144.\vs
\bibitem{juerg}J. Gasser, Nucl. Phys. B279 (1987) 65; J. Gasser,
Proc. 2nd Intern. Workshop on $\pi$N physics (Los Alamos 1987),
eds. W.R. Gibbs and B.M.K. Nefkens, Los Alamos report LA-11184-C 
(1987), p.~266.\vs
\bibitem{glls} J. Gasser, H. Leutwyler, M. Locher and M.E. Sainio,
Phys. Lett. B213 (1988) 85.\vs
\bibitem{MO}M. Olsson, {\tt hep-ph/0001203}, Phys. Lett. B (2000), in print.\vs
\bibitem{JK} J. Kambor, in ``Chiral Dynamics: Theory and Experiment'',
A.M. Bernstein, D. Drechsel and Th. Walcher (eds.) (Springer, Berlin, 1998).\vs
\bibitem{bkmss} V. Bernard, N. Kaiser and Ulf-G. Mei{\ss}ner,
Z. Phys. C60 (1993) 111.\vs
\bibitem{paul}P. B\"uttiker and Ulf-G. Mei{\ss}ner,
Nucl. Phys. A668 (2000) 97.\vs
\end{thebibliography}
\end{document}